\documentclass[a4paper,10pt]{article}

\usepackage[utf8]{inputenc}
\usepackage[english]{babel}
\usepackage[T1]{fontenc}
\usepackage{fullpage}
\usepackage{lmodern}
\usepackage{amssymb}
\usepackage{graphicx}
\usepackage[lofdepth,lotdepth]{subfig} 
\usepackage{amsfonts}
\usepackage{paralist}
\usepackage{float}
\usepackage{tabu}
\usepackage{svg}
\usepackage{amsthm}
\usepackage{float} 
\usepackage{mathtools}
\usepackage{color}
\usepackage{parskip}
\usepackage{authblk}
\usepackage{multirow}
\usepackage{cite}

\usepackage{verbatim}

\newcommand{\1}{{\bf 1}}

\newcommand{\cB}{\mathcal{B}}
\newcommand{\cF}{\mathcal{F}}
\newcommand{\PP}{\mathbb{P}}
\newcommand{\RR}{\mathbb{R}}
\newcommand{\NN}{\mathbb{N}}

\newcommand{\cs}{\mathbf{s}}

\newcommand{\yy}{\mathbf{y}}
\newcommand{\dd}{\mathbf{d}}
\def\permille{\ensuremath{{}^\text{o}\mkern-5mu/\mkern-3mu_\text{oo}}}

\makeatletter
\let\cite\relax
\DeclareRobustCommand{\cite}{%
  \let\new@cite@pre\@gobble
  \@ifnextchar[\new@cite{\@citex[]}}
\def\new@cite[#1]{\@ifnextchar[{\new@citea{#1}}{\@citex[#1]}}
\def\new@citea#1{\def\new@cite@pre{#1}\@citex}
\def\@cite#1#2{[{\new@cite@pre\space#1\if\relax\detokenize{#2}\relax\else, #2\fi}]}
\makeatother

\newtheorem{theorem}{Theorem}[section]

\title{An interaction point process for Bayesian detection of multiple sources in groundwaters from hydrochemical data}

\author[1]{C. Reype}
\author[1]{R. S. Stoica}
\author[2]{A. Richard}
\author[1]{M. Deaconu}

\affil[1]{Universit\'e de Lorraine, CNRS, Inria, IECL, F-54000 Nancy, France}
\affil[2]{Universit\'e de Lorraine, CNRS, GeoRessources, F-54000 Nancy, France}

\begin{document}

\maketitle

\begin{abstract}
Detecting the number and composition of multiple sources in groundwaters from hydrochemical data has remained highly challenging. This paper presents a new interaction point process that integrates geological knowledge for the purpose of automatic sources detection of multiple sources in groundwaters from hydrochemical data. The observations are considered as spatial data, that is, a point cloud in a multidimensional space of hydrochemical parameters. The key assumption of this approach is to consider the unknown sources to be the realisation of a point process. The probability density describing the sources distribution is built in order to take into account the multidimensional nature of the data and specific physical rules. These rules induce a source configuration able to explain the observations. This distribution is achieved with prior knowledge regarding the model parameters distributions. The composition of the sources is estimated by the configuration maximising the joint proposed probability density. The method was first calibrated on synthetic data and then tested on real data from geothermal and ore-forming hydrothermal systems.

\end{abstract}

\section{Introduction}
\label{sec:Introduction}

\begin{figure}[H]
\centering
\includegraphics[scale=0.5]{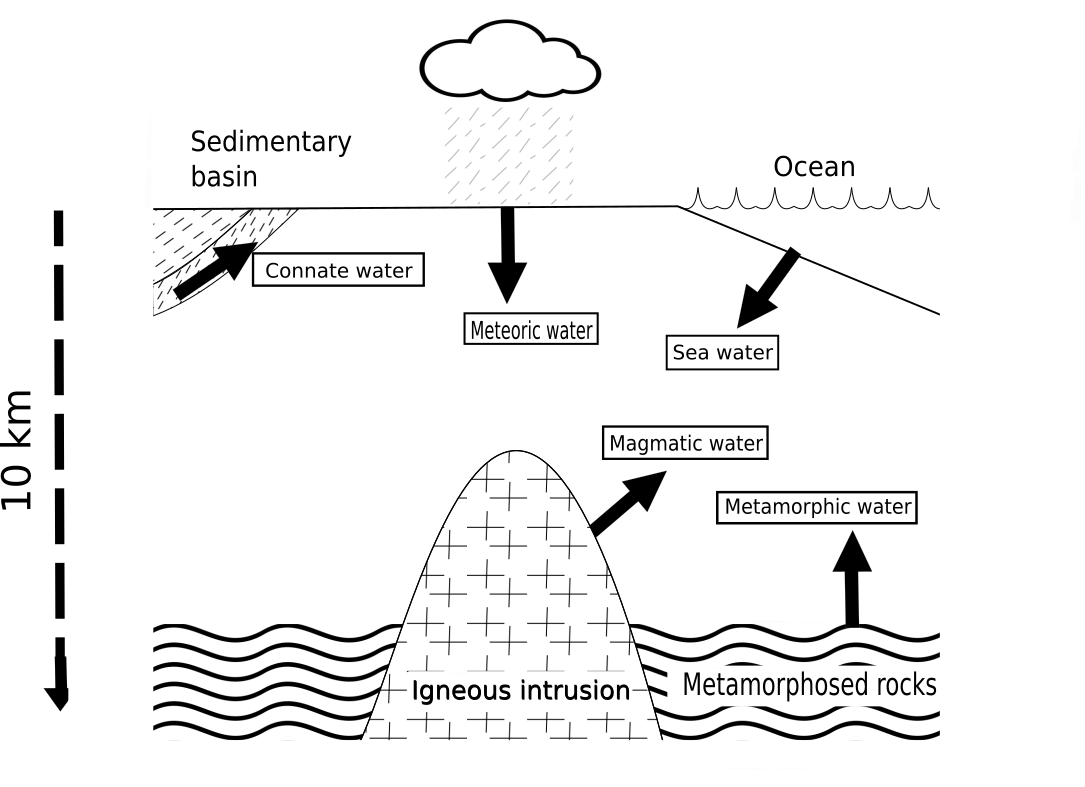}
\caption{Conceptual cross-section of the Earth’s continental crust showing the main sources of surface and deep waters (modified from \cite{Rob05}).}
\label{fig: contexte}
\end{figure}

The analysis of hydrochemical data can be used to build conceptual and quantitative models of fluid and mass transfer in the sub-surface and the Earth's crust \cite{Fau97,YarBod14,IngSan06}. The composition of many groundwaters is controlled by mixing of two or more water sources. The main sources of surface and sub-surface waters which contribute to the composition of groundwaters in hydrothermal systems through mixing processes are shown in the Figure \ref{fig: contexte} (a)). Similar water mixing processes also occur in surface and shallow subsurface environments, potentially involving other water sources. In such cases, the analysis of hydrochemical data includes detecting the sources involved with mixing (\textit{i.e.} number and composition) and estimating their contribution to the data (respectively "inverse analysis" and "forward analysis").

The sources (also refer to as end-members) are mixed together at variable proportions to result in the samples also called mixing terms. The values for each hydrochemical parameter are determined from direct sampling (from boreholes or springs) or from fluid inclusions. Hydrochemical parameters considered are the concentration of ions or molecules, isotopic composition or ratios of hydrochemical parameters. The data are seen as spatial data: a sample is represented by a point in the data space, with coordinates being the value of each hydrochemical parameter. Hence, mentions of position and distance in this paper will not refer to sample location but composition (\textit{e.g.} measurement of all hydrochemical parameters) and difference in composition respectively, whereas mentions of dimension refer to composition in a hydrochemical parameter. In the context of fluid mixing, a sample is considered as a barycentre of the sources: a data point $d_j$ is a result of a mixing between the pattern of sources $\cs=\{s_1,\dots,s_n\}$ if
\begin{equation}
d_j=\sum_{i=1}^n \gamma_{(j);i} s_i,
\label{eq: mixing}
\end{equation}
with $0\leq\gamma_{(j);i}\leq 1$ the contribution of the source $i$ in the point $j$. The Figure \ref{fig: contexte} (b) is an example of a ternary diagram showing typical mixing scenario in the space of hydrochemical parameters (solute1, solute2, solute3). Blue symbols represent the sources/end-members. Black dots represent the samples/mixing terms.
 
If the sources are known, the contribution of the sources can be estimated by using either Bayesian mixing models \cite{LonVerErs18,AreAciHet15,CarVaz04,SkuLon15,ParIng10,PhiGre00,LajRen20,TipSha20} or a likelihood uncertainty estimation that relies on End Member Mixing Analysis (EMMA) \cite{DelEss13}.

The already existing solutions, proposed to solve the problem of sources detection, try to tackle three challenges: the multidimensional nature of the data, the unknown number of sources and physical constraints. The first constraint is to minimise the number of sources. The second constraint is to select sources that explain the data (\textit{i.e.} the convex hull of the sources tends to enclose the data). The third constraint is to consider that the data are representative of the mixing system (\textit{i.e.} the convex hull of the sources is outlined by the data). The last constraint is to consider that the composition of the sources are significantly different from another.

To the best of our knowledge, the existing methods presented in the literature do not take into account all these aspects together. Source detection is done either graphically or in a supervised statistical analysis. A principal component analysis (PCA) is sometimes used to guide the choice of the number of sources \cite{ChrHoo92}. Based on this, the composition of the sources can be estimated by an "end-member mixing analysis" (EMMA) \cite{Wel97}. When the number of sources is known, a more geometrical method can be used: the sources can be estimated by the vertex of the smallest triangle (in terms of area) that contains the data in the case of three sources in a bidimensional data space \cite{Pint20}. 

This paper develops a new Bayesian method of sources detection in hydrochemical data. This procedure is inspired by pattern detection methodologies used in image analysis, animal epidemiology and astronomy \cite{StoDesZer03,StoiGayKret07,StoMaSa05,StoMaSa07}.
It has the advantages to be unsupervised and to take into account simultaneously the previously mentioned physical constraints. Furthermore, the model considers the pattern of sources with no condition on the maximum number of sources. Conditionally to the parameters of the model, the probabilistic source model considered is a Gibbs point process that controls the distribution of the sources in the data space. The set of sources is estimated by the configuration of points that maximises the joint probability density controlling the sources and the parameters distributions. The model presented in this paper is called Hug model in reference to the way that the sources enclose the data. The optimisation procedure is implemented via a simulated annealing procedure, hence avoiding local maximum. The sampling Markov chain Monte Carlo (MCMC) algorithm at the basis of the simulated annealing procedure is achieved by a Metropolis-Hastings within Gibbs sampler. This allows to deal with the multidimensional aspect of the problem.

The conditions for using the Hug model are as follows: the datasets are the results of a conservative mixing (\textit{i.e.} no chemical reaction affects the considered hydrochemical parameters during the mixing process). The composition of the sources are supposed the same or at least not significantly different for each data points. It is noteworthy that the hydrochemical parameters considered should not induce any curvature in the mixing trends projected on binary plots \cite{LanVoc78}. 


The structure of the paper is as follows. Fundamental notions on point processes, their properties and simulation algorithms are presented in Section 2. Section 3 is dedicated to the description of the Hug model. The proposed solutions exhibit two main components. The first component is represented by a Gibbs point process controlling the source distribution. The second component is Metropolis within Gibbs sampler that allows to sample from the model while taking into account the multidimensional nature of the data. Inference procedures are also presented. Section 4 presents the application of the method on synthetic data. The application on the synthetic data allows tuning the parameters priors. The application on real data from hydrothermal systems permits to test and to analyse the method's performances.

\section{Point processes : definition, properties and simulation}
\label{sec:PointProcesses}

\subsection{Point processes}
Let $(S, \cB, \nu)$ be a measure space, where $S$ is a compact subset of $\RR^d$ of strictly positive Lebesgue measure $0 < \nu(S) < \infty$ and $\cB$ the associated Borel $\sigma$-algebra of subsets of $S$. For $n \in \NN$ let $S_n$ be the set of all unordered configurations $\cs = \{s_1,s_2,\ldots,s_n\}$ of $n$ points $s_i \in S$. Let us consider the configuration space $\Omega = \cup_{n=0}^{\infty}S_n$ equipped with the $\sigma$-algebra $\cF$ generated by the mappings
\begin{equation*}
\{s_1,s_2, \ldots, s_n\} \longmapsto \sum_{i=1}^{n} \1_{\{s_i \in B\}}
\end{equation*}
counting the number of points in Borel sets $B \in \cB$. A point process on $S$ is a measurable map from a probability space into $(\Omega,\cF)$. For introductory material on point processes, we refer the reader to the textbooks by \cite{Lies00,MollWaag04}.

Maybe the most known point process is the homogeneous Poisson point process, constructed as it follows. First, the number of points $n$ in a configuration is chosen according to a Poisson law of parameter $\rho\nu(S)$ with $\rho > 0$ a positive constant named intensity. Then, the $n$ points are spread uniformly independently in $S$. We write $X \sim Poisson(S,\rho)$.

The process $Poisson(S,1)$ , or more specifically its measure, is often considered as a reference measure, $\mu$ \textit{i.e.} ($\forall B\in\cB:\, \mu(B)=\rho\nu(B)$), for more elaborate models. For instance, the inhomogeneous Poisson point process driven by the intensity function $\rho : S \rightarrow \RR^{+}$ has the probability measure for all $F\in\cF$~:

\begin{equation*}
    \PP(X \in F)=\sum_{n=0}^{\infty}\frac{\exp[-\nu(S)]}{n!}\int_S \cdots \int_S \1_{F}\{s_1,\cdots,s_n\}\bigg(\prod_{i=1}^n \rho(s_i)\bigg)\,d\nu(s_1)\cdots d\nu(s_n).
\end{equation*}
In this case, the process is spread in $S$ independently according to the probability density $\rho(\cdot)/\int_{S}\rho(s)d\nu(s)$. Clearly, the probability density of this process with respect to the unit intensity stationary Poisson process is given by

\begin{equation*}
 p(\cs) =  \zeta \prod_{i=1}^{n(\cs)} \rho(s_i),
\end{equation*}
with $\zeta = \exp\left[\nu(S) - \int_{S}\rho(s)d\nu(s) \right]$ the normalising constant and $n(\cs)$ the cardinality of $\cs$. The fact that their distribution is entirely known, makes Poisson processes extremely interesting candidates for numerous modelling approaches. Nevertheless, the independence assumption implies that no interactions of points are considered.

Gibbs points processes are models that take into account interactions of points by means of probability density with respect to the reference measure $\mu$. The general form of this probability density is
\begin{equation}
p(\cs) =  \zeta \exp[-U(\cs)],
\label{eq:gibbsDensity}
\end{equation}
with $U(\cs)$ the energy function specifying the points interactions in a  configuration. Still, in this case, the normalising constant $\zeta^{-1} = \int_{\Omega}  \exp[-U(\cs)] d\mu(\cs)$ is no more available in analytical closed form.

There is a lot of freedom in specifying energy functions, provided the resulting probability density integrates to $1$. This is ensured if the model is locally stable, that is there exists $\Lambda \in\RR^{+}$ such that
\begin{equation}
\frac{p(\cs \cup \{\eta\})}{p(\cs)} \leq \Lambda, \quad \forall \cs \in \Omega, \eta \in S.
\label{eq:localStability}
\end{equation}

There exist less restrictive conditions that ensure the integrability of point process~\cite{Ruel99}. The preference for locally stable models~\eqref{eq:localStability} is due to the good convergence properties induced to the corresponding simulation algorithms~\cite{Lies00,MollWaag04}.

\subsection{Simulation}
Sampling from Gibbs point process densities~\eqref{eq:gibbsDensity} is not trivial. This is due to the fact that the normalising constant $\zeta$ is not available in analytically closed form. The adopted solutions within this context are given by Markov chain Monte Carlo (MCMC) strategies. Among them let us mention : spatial birth-and-death processes, perfect sampling methods, Metropolis-Hastings algorithms, etc. The interested reader may refer to~\cite{BaddEtAl16,Lies00,MollWaag04,Lies00,Geye99} for details and thorough mathematical presentations.

The principle behind the MCMC methods is to simulate a Markov chain that has, as equilibrium distribution, the probability distribution of interest. In our case, this is
\begin{equation}
\pi(A)=\int_A p(\cs)\mu(d\cs), A \in \cF.
\end{equation}

The Metropolis-Hastings (MH) algorithm for point processes, implements a Markov chain whose transition kernel is built using three types of moves or transitions: adding a point to the current configuration (birth), deleting a point from the current configuration (death) and changing a point from the current configuration into a new position (change). Let $p_b,p_d,p_c\in [0,1]$, with $p_b+p_d+p_c \leq 1$, be the probability of birth, death and change, respectively. With probability $p_b$ the move birth is selected: a new point $\eta$ is generated according to a distribution $b(\cs,\eta)$. With probability $p_d$ the move death is selected: a point $\eta$ chosen in the configuration according to a distribution $d(\cs,\eta)$ is deleted. With probability $p_c$ the move change is selected: a point $\eta$ chosen in the configuration according to a distribution $q(\cs,\eta)$ is changed into a location $\zeta$ generated according to a distribution $c(\cs,\eta,\zeta)$. As indicated in the frame below, this transition kernel is embedded within an Update procedure that is to be iterated in order to obtain the desired samples.

\fbox{
\begin{minipage}{\textwidth}
\begin{enumerate}
\item[] \textbf{Algorithm MH~:} $\yy$ = Update ($\cs$)
\begin{enumerate}
\item[1)] Choose a transition type according to $p_b$, $p_d$ and $p_c$, such that $p_b+p_d+p_c \leq 1$.
\item[2)] If a ``birth'' is chosen, generate a new point $\eta$
according to $b(\cs,\eta)$. Accept the new configuration $\yy = \cs
\cup \{\eta\}$ with probability 
\begin{equation*}
\alpha(\cs,\cs \cup \{\eta\}) = \min\{1,r(\cs,\eta)\},
\label{acceptBirth}
\end{equation*}
with 
\begin{equation}
r(\cs,\eta) = \frac{p_d d(\cs \cup \{\eta\},\eta)}{p_b b(\cs,\eta)}\frac{p(\cs \cup \{\eta\})}{p(\cs)}.
\label{acceptance_ratio}
\end{equation}
\item[3)] If a ``death'' is chosen, select a candidate $\eta$ to be
deleted from $\cs$ according to $d(\cs,\eta)$. Accept the new
configuration $\yy = \cs \setminus \{\eta\}$ with probability
\begin{equation*}
\alpha(\cs,\cs \setminus \{\eta\}) = \min\{1,1/r(\cs \setminus \{\eta\},\eta)\}.
\label{acceptDeath}
\end{equation*}
\item[4)] If a ``change'' is chosen, select a candidate $\eta$ from $\cs$ according to $q(\cs,\eta)$ and change it into a new candidate $\zeta$ according to $c(\cs,\eta,\zeta)$. Accept the new
configuration $\yy = \cs \setminus \{\eta\}\cup \{\zeta\}$ with probability
\begin{equation*}
\alpha(\cs,\cs \setminus \{\eta\}\cup \{\zeta\}) = \min\{1,\min\{1,r(\cs,\eta,\zeta)\}\},
\label{acceptChange}
\end{equation*}
with 
\begin{equation}
r(\cs,\eta,\zeta) = \frac{q(\cs \setminus \{\eta\}\cup \{\zeta\},\eta)}{q(\cs,\eta)}\frac{ c(\cs \setminus \{\eta\}\cup \{\zeta\},\zeta,\eta)}{ c(\cs,\eta,\zeta)}\frac{p(\cs \setminus \{\eta\}\cup \{\zeta\})}{p(\cs)}.
\label{acceptance_ratio_c}
\end{equation}
\end{enumerate}
\end{enumerate}
\end{minipage}
}


Maybe the most adopted birth and death proposals are the uniform choices
\begin{equation*}
b(\cs,\eta) = \frac{\1_{\{\eta \in S\}}}{\nu(S)}
\end{equation*}
and
\begin{equation*}
d(\cs,\eta) = \frac{\1_{\{\eta \in \cs\}}}{n(\cs)}, 
\end{equation*}
respectively. The uniform choice is also adopted for the event change. Hence $q(\cs,\eta)=d(\cs,\eta)$ and the new point is generated in the ball centred in $\eta$ and of radius $r_c\in\RR^{+}$ noted $B(\eta, r_c)$:
\begin{equation*}
c(\cs,\eta,\zeta)= \frac{\1_{\{\zeta \in B(\eta, r_c)\}}}{B(\eta, r_c)}.
\end{equation*}


These choices together with the local stability~\eqref{eq:localStability} guarantee the geometric ergodicity, Harris recurrence and $\phi-$irreducibility of the Markov chain simulated using this Metropolis-Hastings algorithm~\cite{Geye99,Lies00,MollWaag04}. For our situation, this implies that the simulated Markov chain with this transition kernel converges towards the distribution of the proposed model from any initial condition, with a geometric speed.

\subsection{Inference}
In the following, we assume that we are in the possession of a well-defined source model $p(\cs)$ which is a point process density and of an appropriate sampling algorithm able to sample from it. 

Under these circumstances, maximisation of the probability density can be achieved via a simulated annealing algorithm based on the previously described MH dynamics. This algorithm iteratively draws samples at a temperature $T\in\RR^{+}$ from $p(\cs)^{1/T}$ while $T \rightarrow 0$. A logarithmic cooling schedule for the temperature $T$ guarantees the convergence of the simulated annealing towards the uniform distribution of the configurations subspace that maximises $p(\cs)$~\cite{StoiGregMate05}.

The solution to the optimisation problem is not guaranteed by a unique configuration of points. Hence, in order to get more robust results, averaging may be useful. This can be achieved through the computation of level sets.  

Let $X$ be a point process and $S = \bigcup_{i=1}^{m}\tilde{s_i}$ a decomposition of the domain $S$ in a finite number of cells $m$, such that $\nu(\tilde{s_i}) = \text{ct.}$, for all $i$. Let $\tilde{S}$ be the set of all the cells. The contact probability  between the point process and a grid cell is 
\begin{equation*}
p(s) = \PP(s \cap X \neq \emptyset).
\end{equation*}
For $\lambda \in [0,1]$, let us consider the level set given by
\begin{equation*}
l_{\lambda} = \{ s \in \tilde{S}~: p(s) > \lambda \}.
\end{equation*}
The practical computation of $p(s)$ is done via Monte-Carlo methods,  the estimator
\begin{equation*}
p_{n}(s)=\frac{1}{n}\sum_{i=1}^n\1_{\{s \cap S_i \neq \emptyset\}}.
\end{equation*}
with $S_1,S_2,\ldots,S_n$ i.i.d realisations of $p(\cs|\theta)$. Hence, the estimator of the level set is
\begin{equation}
l_{n,\lambda} = \{ w \in \tilde{S}~: p_{n}(s) > \lambda \}.
\label{eq:EstimatorLevelSets}
\end{equation}

Clearly, the sets $l(\lambda)$ are quantiles of the random set $X$. If the random set $X$ is the sources configuration governed by the model $p(\cs)$ the estimators of the level sets may indicate the regions in $S$ that are visited by the model with a probability higher than $\lambda$. The derivation and the properties of these level sets estimators are given in~\cite{HeinEtAl12}.

\section{The Hug model}
\label{sec:Hug}
The data considered are the measurements of $K$ hydrochemical parameters of $m$ samples denoted $\dd=\{d_1,d_2,\ldots,d_m\}$. For instance such a hydrochemical parameter may be the concentration of a given chemical element. The sample or data point numbered $j$ ($1\leq j\leq m$) is placed in the data space with the coordinates $d_j=(d_{(j);1},\dots,d_{(j);K})$. Here $d_{(j);k}\in\RR$ represents the measurement of the hydrochemical parameter numbered $k$ ($1\leq k\leq K$) associate to sample $j$. Hence, the dataset is a point cloud made of $m$ points (or samples) in a $K$-dimensional space of finite volume.

In this space, the pattern of sources is unknown. Still, it is related to the set of the data points. The different aspects of the relationship between the data and the unknown sources can be synthesised by the following assumptions:
\begin{itemize}
\item[(a)] the data points originating from a mixture of sources should be rather close to the sources
\item[(b)] the data points are enclosed within the convex hull given by the positions of the source: this is due to the fact that the data points are barycentres of the sources
\item[(c)] the number of sources is not known but it should be controlled or minimised in a certain sense
\item[(d)] the composition of the sources, that is their location in the data space should be significantly different, from one source to another.
\end{itemize}

By hypothesis (a) the position of the sources are in a bounded space: without it, every dataset can be explained by three sources placed in infinity. The hypothesis (b) is a physical consequence of mixing with mass conservation \cite{Fau97}. Hypothesis (c) and (d) take into account the fact that in practice, the number of sources involved in a mixing is supposed to be less than $10$ \cite{Fau97,YarBod14}. 

The key idea of our work is to build a Gibbs point processes that governs the sources distribution in the data space. The energy function of the process integrates the previous hypotheses. Since, these assumptions specify relations between the data and the unknown sources, but also interactions among sources only (\textit{i.e.} interaction between points of a pattern of points), the probability density of the pattern of sources $\cs$ can be written as follows

\begin{equation}
p_{\dd}(\cs|\theta)=\frac{\exp\left[-U(\cs|\theta)\right]}{Z(\theta)} =\frac{\exp\left[-U_{\dd}(\cs|\theta) - U_{i}(\cs|\theta)\right]}{Z(\theta)}.
\label{eq: model}
\end{equation}

Here $U_{\dd}(\cs|\theta)$ is the data term. It locates the sources in the data space while taking into account the hypotheses (a) and (b). It depends on the observed data $\dd$.
The term $U_{i}(\cs|\theta)$ is the interaction energy that manages the sources relative position by taking into consideration the hypotheses (c) and (d). This term does not depend on $\dd$. Finally, $Z(\theta)$ represents the normalising constant. The sum of the data and interaction terms gives the total energy function $U$.


In the following, we specify the model~\eqref{eq: model}. First, the model is presented when $K=2$, that is the hydrochemical space is a finite surface. Then the model is generalised for $K \geq 2$.

\subsection{Data energy function}
The data term $U_{\dd}(\cs|\theta)$ controls the positioning of the sources with respect to the observed data points. This term allows the model to detect source patterns while taking into account hypotheses (a) and (b).

Having in mind (a), let us consider the ratio between the area of the convex hull of the sources $g(\cs)$ and the area of the convex hull of the data $g(\dd)$. More specifically we consider the statistic $g(\cs,\dd)$:

\begin{equation}
 g(\cs , \dd)=\left\lvert\frac{g(\cs)}{g(\dd)}-1 \right\rvert
\label{eq:gstat}
\end{equation}
with $|\cdot|$ the absolute value function.

If the data and the convex hull determined by the sources tend to have equal surfaces, the statistics value should be close to $0$. Furthermore, $g(\cs,\dd)$ is bounded, since the observation domain is bounded. The numerical computation of~\eqref{eq:gstat} can be performed via the Andrew's monotone chain convex hull algorithm \cite{And79}. 

A simple solution to prevent pathological cases is to require a number of minimum three sources. As it will be shown later, the impact of this supplementary condition is attenuated by the use of level sets and the sequential $k-$means algorithm presented in Section \ref{reconstruction}.


The hypothesis (b) is considered by the following statistic
\begin{equation}
n_e(\cs , \dd)=1-\frac{n_{expl}(\cs,\dd)}{m}
\label{eq:nestat}
\end{equation}
where $n_{expl}(\cs,\dd)$ is the number of points explained by the pattern of sources, (\textit{i.e.} the number of points of $\dd$ inside the convex hull of $\cs$) and $m$ the total number of samples. Whenever the sources tend to explain all the points, the statistic~\eqref{eq:nestat} is close to $0$.

The data energy function is:
\begin{equation}
U_{\dd}(\cs|\theta) = \theta_1 g(\cs,\dd) +  \theta_2 n_e(\cs,\dd) ,
\label{eq:dataenergy}
\end{equation}
with $\theta_1,\theta_2 \in \RR^{+}$ the model parameters controlling the strength of each statistic and so the weight of hypothesis (a) and (b) respectively.

\subsection{Interaction energy function}
The term $U_{i}(\cs|\theta)$ controls the sources interactions and it does not depend on the data $\dd$. This term allows taking into account the hypotheses (c) and (d). 

The number of sources $n(\cs)$ in a configuration controls the (c) hypothesis, while the proximity of sources required by the hypothesis (d) is controlled by $n_{r}(\cs)$. This last statistic represents the number of pairs of sources situated within a pre-fixed distance $r$ from each other:
\begin{equation}
   n_{r}(\cs)=\sum_{i=1}^n \sum_{j=i+1}^n \1\{d(s_i,s_j)\leq r\},
\end{equation}
here $d(s_i,s_j)$ is the Euclidean distance between $s_i$ and $s_j$.

The interaction energy function is:
\begin{equation}
U_{i}(\cs|\theta) = \theta_3 n(\cs)+\theta_4 n_{r}(\cs) ,
\label{eq:interactionenergy}
\end{equation}
with $\theta_3,\theta_4 \in \RR^{+}$ the model parameters controlling the weight of hypothesis (c) and (d) respectively.

\subsection{Source estimator}
The Hug model is a Gibbs point process defined by the energy functions~\eqref{eq:dataenergy} and~\eqref{eq:interactionenergy}.

Assuming knowledge related to the model parameters is available through the prior $p(\theta)$, the joint distribution is written as
\begin{equation*}
p_{\dd}(\cs,\theta)=p_{\dd}(\cs|\theta)p(\theta). 
\label{eq:jointdistrib}
\end{equation*}

Within this context, the unknown source pattern together with its parameters are estimated by maximising~\eqref{eq:jointdistrib}:
\begin{equation}
\widehat{(\cs,\theta)} = \arg\max_{\Omega \times \Theta} 
p_{\dd}(\cs,\theta) = \arg\max_{\Omega \times \Theta} p_{\dd}(\cs|\theta)p(\theta)
\label{eq:jointestimator}
\end{equation}
with the configuration space $\Omega$ and the parameter space $\Theta$ a compact region in $\RR^4$. The computation of~\eqref{eq:jointestimator} can be achieved via a simulated annealing algorithm. 

\subsection{General case: $K \geq 2$}
The observed datasets contain a number of hydrochemical parameters greater than two. The construction of a solution in dimension $K$ by the generalisation of~\eqref{eq:jointestimator} is mathematically possible. Still, this straightforward approach may lead to extremely heavy computations.

Here, an alternative solution is preferred. Assuming the dataset contains $K$ hydrochemical parameters, let us consider all the planes obtained by taking pairs of hydrochemical parameters. The total number of different planes is $L=K(K-1)/2$ (considering hydrochemical parameter $i_1$ and $i_2$ is the same as considering $i_2$ and $i_1$). The solution we propose is to achieve the distribution $p_{\dd}(\cs,\theta)$ with an auxiliary variable that selects such a plane, hence allowing operations only in spaces of dimension two. This auxiliary variable is discrete and finite taking values in $V = \{1, 2,\ldots,L\}$. It is governed by the prior distribution $p(v)$.\\

Let us consider the following model
\begin{equation*}
p_{\dd}(\cs,\theta,v)=p_{\dd}(\cs,\theta|v)p(v) = p_{\dd}(\cs|\theta,v)p(\theta)p(v). 
\end{equation*}

The conditional distribution is given by
\begin{equation}
p_{\dd}(\cs|\theta,v) \propto \exp[- U(\cs|\theta,v)],
\label{eq:totalHug}
\end{equation}
with energy function
\begin{equation*}
U(\cs|\theta,v) = U_{\dd}(\cs|\theta,v) + U_{i}(\cs|\theta,v).
\end{equation*}

The data energy expression writes
\begin{equation*}
U_{\dd}(\cs|\theta,v) = \sum_{l = 1}^{L} U_{\dd}^{(l)}(x|\theta) \1_{\{v=l\}}
\end{equation*}
where each element in the sum is 
\begin{equation}
U_{\dd}^{(l)}(x|\theta) = \theta_1 g^{(l)}(\cs,\dd) +  \theta_2 n_e^{(l)}(\cs,\dd), \quad l = 1,\ldots,L,
\end{equation}
with $g^{(l)}(\cs,\dd)$ and $n_e^{(l)}(\cs,\dd)$ the data energy statistics corresponding to the parametric plane numbered~$l$.

The energy interaction is developed in analogous manner:
\begin{equation*}
U_{i}(\cs|\theta,v) = \sum_{l = 1}^{L} U_{i}^{(l)}(x|\theta) \1_{\{v=l\}}
\end{equation*}
where each element in the sum is 
\begin{equation}
U_{i}^{(l)}(x|\theta) = \theta_3 n(\cs)+\theta_4 n_{r}^{(l)}(\cs), \quad l = 1,\ldots,L,
\end{equation}
with $n_r^{(l)}(\cs,\dd)$ the number of interacting pairs of sources corresponding to the plane numbered~$l$.

This framework allows proposing as joint estimator for the source pattern, model parameters and planes selector
\begin{equation}
\widehat{(\cs,\theta,v)} = \arg\max_{\Omega \times \Theta \times V} 
p_{\dd}(\cs,\theta,v) = \arg\max_{\Omega \times \Theta \times V} p_{\dd}(\cs|\theta,v)p(\theta)p(v).
\label{eq:totaljointestimator}
\end{equation}

The proposed construction is a mixture of bidimensional processes, as in the graphical detection made by hydrogeologist. Furthermore, as it will be shown just below, the sampling of $p_{\dd}(s,\theta,v)$ can be done using a Metropolis-Hasting within Gibbs algorithm, allowing the computation of~\eqref{eq:totaljointestimator} based on a simulated annealing dynamics.


\subsection{Simulation of the Hug model and implementation of the simulated annealing algorithm}

\begin{theorem}
Let us assume for all $l\in[1,L]$ that $g^{(l)}(\dd)>0$. The HUG model with $K=2$ is integrable for all $\theta_1, \theta_2,\theta_3,\theta_4 > 0$.
\end{theorem}

The proof of this theorem is presented in the Appendix \ref{secA1}.


The previous result allows to sample the HUG model with $K=2$ with the MH algorithm presented in Section 2.

There is no information available regarding the convexity of $p_{\dd}(\cs,\theta,v)$. The computation of~\eqref{eq:totaljointestimator} requires a global optimisation procedure. For this purpose, a simulated annealing algorithm may be implemented. 

Here, the simulated annealing samples iteratively from $p(\cs,\theta,v)^{1/T}$ while the temperature $T$ goes slowly to $0$. 

The sampling from $p(\cs,\theta,v)$ a MH within Gibbs dynamics. The priors $p(\theta)$ and $p(v)$ are chosen to be easy to sample. Once a parameter and a hydrochemical plane are chosen by their priors, respectively, sampling from $p(\cs|\theta,v)$ is just the simulation of bidimensional Hug model. This step is performed using the MH algorithm in Section 2.

Regarding the cooling schedule, the authors in~\cite{StoiGregMate05} proved that a logarithmic scheme guarantees the convergence of the simulated annealing based on MH algorithms for point processes. For speeding up the computation time, here preference is given to the polynomial scheme
\begin{equation*}
T_{n+1} = c T_{n}, \quad c \in ]0,1[.
\end{equation*}

The general algorithm is:

\fbox{
\begin{minipage}{\textwidth}
\begin{enumerate}
\item[]\textbf{Algorithm SA~:} Fix $p(\theta)$ and $p(v)$. Choose a random initial configuration $\cs_0$. The initial temperature is $T_1$, the cooling coefficient is $c$, the total number of iterations is $N$, $G$ the number of applications of the Gibbs sampler and $M$ the number of applications of the MH algorithm.
\item For k=1 to N do
\begin{itemize}
\item $\theta_k \sim [p(\theta)]^{1/T_k}$
\item for g=1 to G do
    \begin{enumerate}
        \item $v_k \sim [p(v)]^{1/T_k}$
        \item $\cs_{k} \sim [p(\cs|\theta_k,v_k)]^{1/T_k}$. This step is achieved by calling MH$(\cs_{k-1},T_{k})$ successively $M$ times.
    \end{enumerate}
\item $T_{k+1} = c T_{k}$
\end{itemize}
\item Return $(\cs_{N},\theta_{N},v_{N})$.
\end{enumerate}
\end{minipage}
}

\section{Application}
\label{sec:Application}
This section demonstrates the proposed method application. First, the normalisation of the dataset and the parameter set-up for each of the presented algorithms are explained. Then the  model is applied on synthetic datasets. The first synthetic dataset allows us to evaluate the results of the model when the real sources are known and visible on each projection plane. The second synthetic dataset accounts for a mixing of four sources and considers three hydrochemical parameters. The sources are positioned such as only $3$ sources are visible on each projection plane. Finally, the Hug model is then applied on real datasets.

\subsection{Data and parameters set-up}
\label{transformation}
For all the datasets, a normalisation procedure is built such that the data and the simulated sources are in the unit hyper-cube $W=[0.0,1.0]^K$. The normalisation is made dimension by dimension. More precisely, for each dimension we define the window of the range of values that a source can take and, by a linear transformation, convert it into $[0.0,1.0]$. This range is set for each dimension $k\in[1,\ldots, K]$ to $(\min_{j}(d_{(j),k})-\delta_k , \max_{j}(d_{(j),k})+\delta_k)$ where $\delta_k$ is a threshold set by the user. Here we take $\delta_k=\max_{j}(d_{(j),k})-\min_{j}(d_{(j),k})$. Regarding the interaction radius needed by the Hug model, the value $r=0.01$ is chosen for each projection plane. If available, the Bayesian framework allows integrating prior knowledge regarding the threshold values $\delta_k$ and the radius $r$.

The algorithms parameters were chosen based on trial and error procedures. The SA algorithm was run for $N=3.5*10^6$ iterations. For each iteration, the Gibbs dynamics was applied $G=L$ times.
Each time the MH is called, it performs $M=200$ steps. The initial temperature is $T_1=2*10^4$ and the cooling coefficient is $c=0.99999$. The temperature is cooled until $T_{min}=10^{-6}$. The last $10^6$ iterations are performed at constant temperature. At this very low temperature, these last outputs of the algorithms may be considered closed enough to the desired solution~\eqref{eq:totaljointestimator}. Furthermore, they tend to be identically distributed, allowing the computation of level sets and of robust statistics.

The probabilities selecting the possible transitions allowed by the MH kernel were fixed as follows: $p_b=0.2$ for “birth”, $p_d=0.2$ for “death” and $p_c=0.6$ for “change”. The support of the uniform proposal in “change” is given by $r_c=0.3$.

The Table~\ref{tab: init} gives a synthetic presentation of the previously mentioned variables.

\begin{table}[H]

 \begin{tabular}{|*{3}{c|}} 
 \hline 
Variable & Description & Value  \\ 
 \hline
$L$ & number of planes & $K(K-1)/2$\\
$\delta_k$ & threshold of observation & $\max_{j}(d_{(j),k})-\min_{j}(d_{(j),k})$\\
$r$ & interaction radius & $0.01$\\
$N$ & SA iterations & $3.5*10^6$\\
$G$ & number of applications of the Gibbs sampler & $L$\\
$M$ & number of steps in the MH algorithm & $200$\\
$T_1$ & initial temperature & $10^4$\\
$c$ & cooling coefficient & $0.99999$\\
$p_b;p_d;p_c$ & probabilities of birth;death;change & $0.2;0.2;0.6$\\
$r_c$ & support of the uniform “change” proposal & $0.3$\\
 \hline 
\end{tabular}
\caption{Data normalisation, model interaction radius and algorithms parameters.}
\label{tab: init}
\end{table}

The initial configuration of sources is given by distributing $4$ points uniformly in $W$. The algorithm outputs, that is the pattern of sources, its statistics and the parameters $\theta$ are saved every $1000$ iterations. This gives a total of $1000$ saved samples containing the detected patterns of sources and their associate sufficient statistics. Only the last $500$ saved patterns are used for statistical inference.

For the prior $p(v)$ the uniform distribution was adopted. The prior $p(\theta)$ was chosen following a strategy similar to classical Approximate Bayesian Computation (ABC) principles \cite{Blu10}. The Hug model with different pre-fixed $\theta$  values was applied on several synthetic datasets with known sources. The parameters providing sources close to the real sources were kept. Hence, the prior $p(\theta)$ was set as a Gaussian distribution. Its parameters were chosen according to the empirical mean and variance of the kept $\theta$ parameters. Table~\ref{tab: theta} shows these prior parameters.

\begin{table}[H]
 \begin{tabular}{|*{5}{c|}} 
 \hline 
 & $\theta_1$  & $\theta_2$  & $\theta_3$  & $\theta_4$  \\ 
 \hline
$\mu_{\theta_i}$ & 11.25 & 250.0 & 0.25 & 1.0 \\ 
 \hline 
$\sigma_{\theta_i}^2$ & 1.0 & 10.0 & 0.01 & 0.01  \\ 
 \hline 
\end{tabular}
\caption{Parameter of the Gaussian prior of $\theta$.}
\label{tab: theta}
\end{table}

\subsection{Synthetic datasets}
In synthetic datasets, the detected sources can be compared to the real known ones. Synthetic datasets are made by first setting the number of dimensions $K$, the number of sources $n$, their positions $\cs^*$ and the number of samples $m$. The position of each sample is created by generating a vector of sources contributions. Here, we assume that a sample or a data point is generated uniformly in the convex hull of the sources. The Dirichlet distribution with parameters $1_n=(1,\dots,1)\in \RR^n$ generates uniformly in $[0,1]^n$ vectors such as the sum of their coordinates equals $1$.

The first synthetic dataset contains $m=200$ points resulting from the mixing system described in the Table~\ref{tab: source sim}. The data space is made by three hydrochemical parameters named “solute1”, “solute2” and “solute3”. Because the data are normalised, these hydrochemical parameters can represent for example either concentration, elemental ratio or isotopic composition, provided no curvature effect occurs. The proposed sources will be updated in $L=3*2/2=3$ planes. 

\begin{table}[H]
    \begin{tabular}{|c||c|c|c|c|}
    \hline
        \textbf{Sources} & solute1 & solute2 & solute3 \\
        \hline
        1 & 0.3 & 0.78 & 0.8 \\
        \hline
        2 & 0.8 & 0.13 & 0.8 \\
        \hline
        3 & 0.7 & 0.7  & 0.1 \\
        \hline
        4 & 0.2 & 0.2  & 0.2 \\
        \hline
    \end{tabular}
    \caption{Position of the real sources ($\cs^*$) for the first synthetic dataset.}
\label{tab: source sim}
\end{table}

For each plane, the Hug model statistics for the known sources are given in Table \ref{tab: true stat syn}.

\begin{table}[H]
 \begin{tabular}{|*{5}{c|}} 
 \hline 
$g(\cs^*,\dd)$ & $n_e(\cs^*,\dd)$ & $n(\cs^*)$ & $n_r(\cs^*)$ & plane\\
 \hline 
0.358501 &0 &4 &0 &1\\
 \hline 
0.294945 &0 &4 &0 &2\\
 \hline 
0.299012 &0 &4 &0 &3\\
 \hline 
\end{tabular}
\caption{Statistics of the real sources for the first synthetic dataset.}
\label{tab: true stat syn}
\end{table}

\begin{figure}[H]
\centering 
\subfloat[]{\includegraphics[scale=0.2]{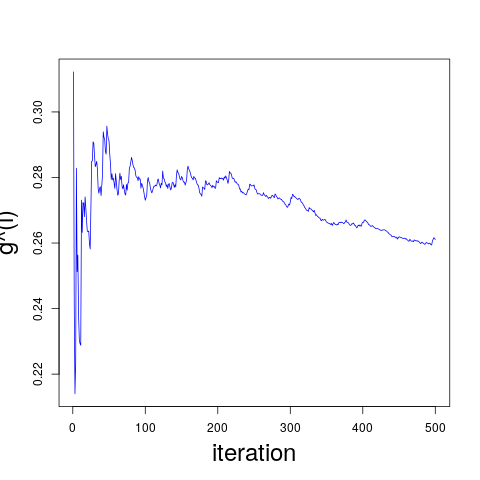}}
\subfloat[]{\includegraphics[scale=0.2]{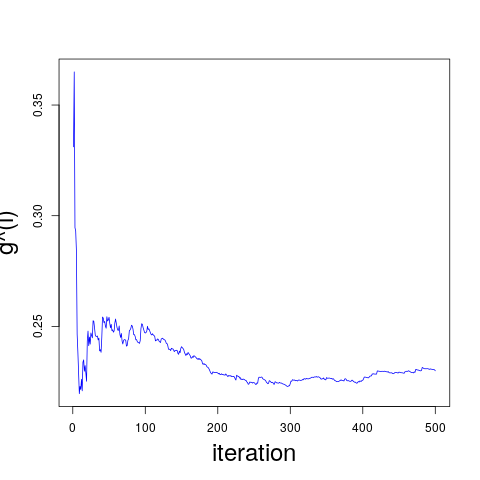}}
\subfloat[]{\includegraphics[scale=0.2]{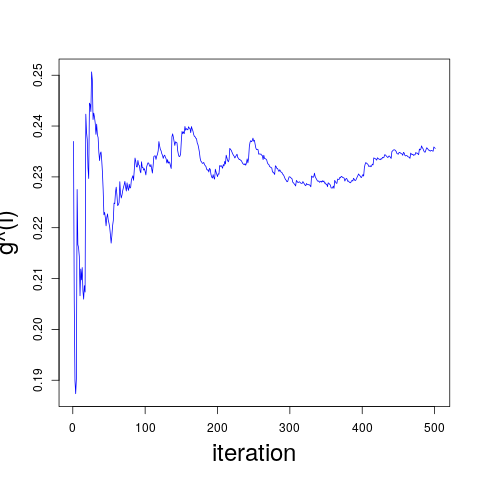}}\\
\subfloat[]{\includegraphics[scale=0.2]{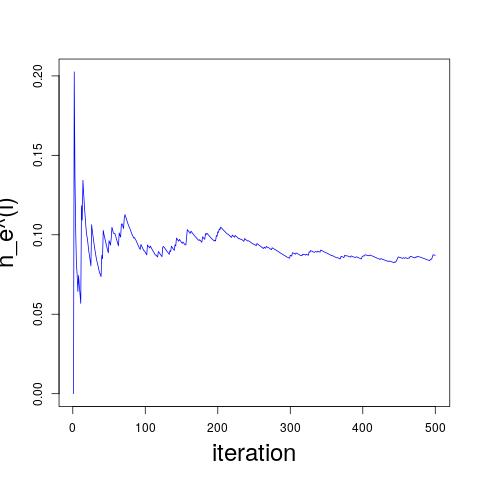}}
\subfloat[]{\includegraphics[scale=0.2]{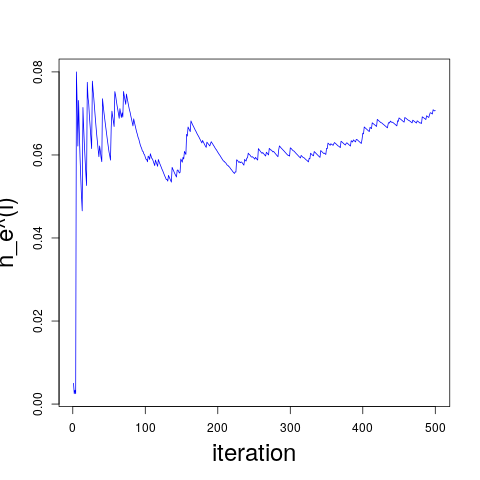}}
\subfloat[]{\includegraphics[scale=0.2]{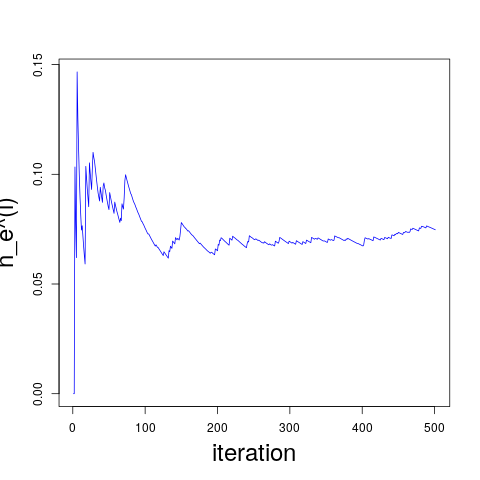}}\\
\subfloat[]{\includegraphics[scale=0.2]{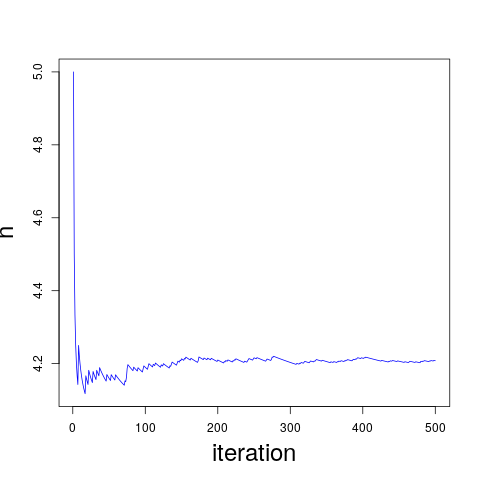}}\\
\subfloat[]{\includegraphics[scale=0.2]{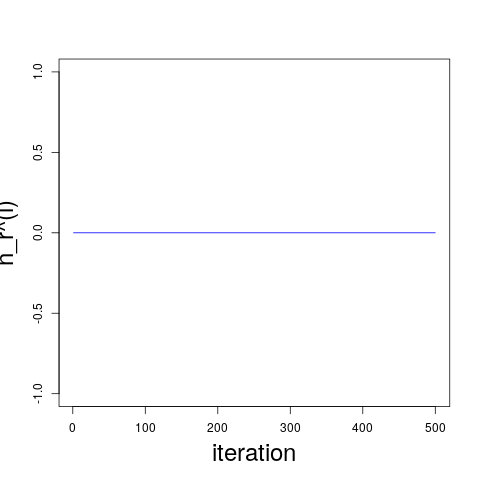}}
\subfloat[]{\includegraphics[scale=0.2]{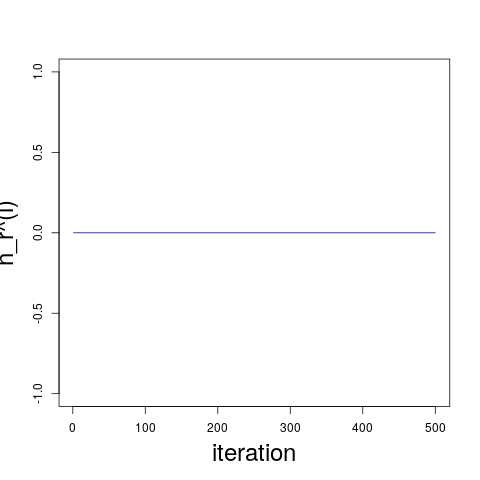}}
\subfloat[]{\includegraphics[scale=0.2]{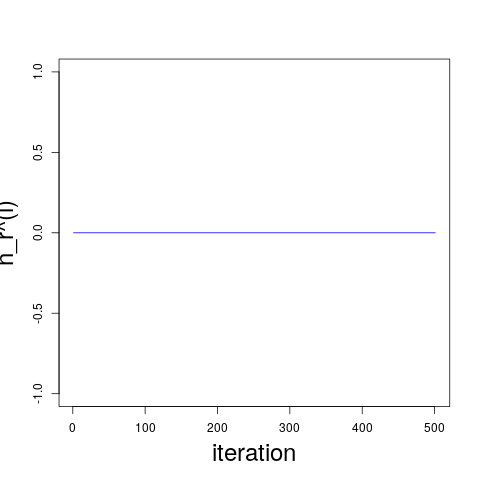}}
\caption{Cumulative means of the statistics for the first synthetic dataset: first plane (a,d,g,h), second plane (b,e,g,i) and third plane (c,f,g,j). Each row represents a statistic.} 
\label{plot: stat sim} 
\end{figure}

The Figure~\ref{plot: stat sim}  shows the cumulative means of the statistics series. It can be observed that they clearly tend to approach constant close to the true statistics obtained from the known sources.

The patterns of the detected sources are projected on every plane. Following~\cite{HeinEtAl12}, level sets are estimated. The computation was done for a regular grid with cells of length $0.02$. The probability of having a detected source in a cell is indicated by the coloured scale. This probability is estimated by~\eqref{eq:EstimatorLevelSets}.

\begin{figure}[H]
\centering 
\subfloat[first plane]{\includegraphics[scale=0.35]{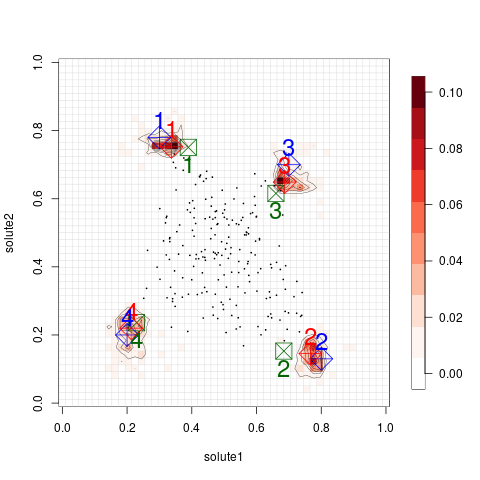}}
\subfloat[second plane]{\includegraphics[scale=0.35]{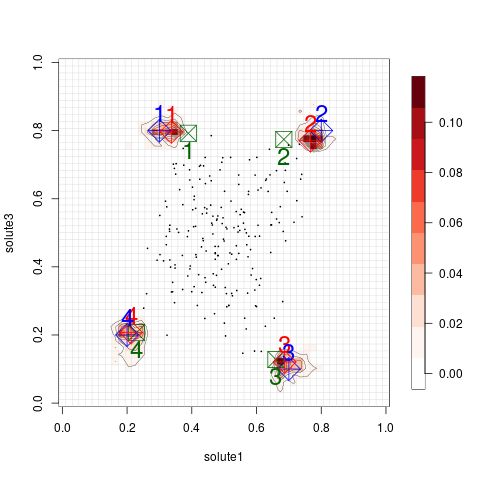}}\\
\subfloat[third plane]{\includegraphics[scale=0.35]{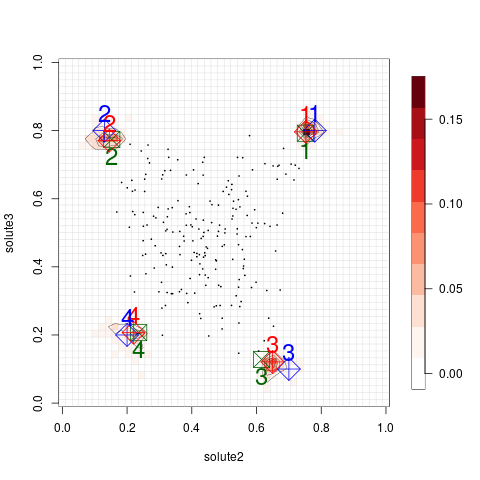}}
\caption{Level sets computed the first synthetic dataset. The blue symbols represent the known sources, the green symbols are the centres of the clusters obtained by the $k$-means algorithm with four clusters on the pattern of sources, and the red symbols are the median points.} 
\label{plot: count sim} 
\end{figure}

The Figure~\ref{plot: count sim} presents the obtained results. The model finds in each plane $4$ regions associated with level sets exhibiting high probability values that indicate the presence of sources. On each plane, these regions are close to the real sources (in blue). These results match the behaviour of the statistics in Figure \ref{plot: stat sim}. The cumulative means of the statistics approach the corresponding values computed from the known sources. The detected sources are grouped in $4$ clusters using a $k$-means algorithm. The position of clusters centres are shown in green in Figure~\ref{plot: count sim}. The median point of each cluster (represented in red) is made by considering the median coordinate of each cluster. The proposed pattern is the pattern of median points specified in the Table~\ref{tab: proposed sources sim}. This choice is adopted in order to diminish the impact of extreme values.

\begin{table}[H]
    \begin{tabular}{|c||c|c|c|c|c|c|c|c|c|c|}
    \hline
       \multirow{2}{*}{\textbf{Sources}} & \multicolumn{3}{|c|}{solute1} & \multicolumn{3}{|c|}{solute2} & \multicolumn{3}{|c|}{solute3} \\
        \cline{2-10}
        & median & mean & sd& median & mean & sd& median & mean & sd\\
        \hline\hline
        1 & 0.69& 0.7& 0.03& 0.64& 0.65 &0.03 &0.11& 0.11& 0.02 \\
        \hline
        2 & 0.77& 0.77 &0.02& 0.13& 0.13 &0.03& 0.76& 0.77& 0.02 \\
        \hline
        3 & 0.21& 0.21&0.02& 0.22& 0.21& 0.04 &0.21& 0.2 &0.03 \\
        \hline
        4 & 0.33& 0.33& 0.03& 0.76& 0.76& 0.02& 0.8& 0.8& 0.02 \\
        \hline
    \end{tabular}
    \caption{Proposed pattern for the first synthetic dataset made by the median points of clusters computed using a $k$-means with $4$ classes on the whole space.}
\label{tab: proposed sources sim}
\end{table}

The distance between the proposed sources $i$ and the real sources $j$ is calculated on the dimension $k$ by the formula $\frac{|s_{(i);k}-s^*_{(j);k}|}{s^*_{(i);k}}\times 100$ which is the relative difference. For each dimension, the average error is given (in $\%$) by “Mean Error Dimension” and the average error for each source is given (in $\%$) by “Mean Error Source” in the Table \ref{tab: error sim}.

\begin{table}[H]
    \begin{tabular}{|c||c|c|c||c|}
\hline
\textbf{Sources} & solute1 & solute2 & solute3 & Mean Error Source\\
\hline
1 &13.3&   3.8   & 0.0 & 5.7\\
\hline
2 & 3.8 &  15.4  & 3.8 & 7.7\\
\hline
3 &1.4 & 7.1 &  20.0 & 9.5\\
\hline
4 & 5.0 & 10.0  &  5.0 & 6.7\\
\hline\hline
Mean Error Dimension&5.9   &  9.1  &   7.2  &  7.4\\
        \hline
    \end{tabular}
    \caption{Relative difference (in $\%$) for the proposed sources with respect to the known ones, and the mean error for each source and each dimension, for the first synthetic dataset.}
\label{tab: error sim}
\end{table}




\subsubsection{Estimation of the source position}
\label{reconstruction}
Due to projection effects that depend on the data structure, detecting the number of sources is not a trivial task. The projected real sources are not always located in the convex hull of the real sources in every plane. The second synthetic dataset is such an example. The dataset contains $100$ points obtained by a mixing system with $4$ sources in a $3$ dimensional space (see Table~\ref{tab: sources tetra}). The special feature of this dataset is that only $3$ are visible on each plane. 

\begin{table}[H]
    \begin{tabular}{|c||c|c|c|c|}
    \hline
        \textbf{Sources} & solute1 & solute2 & solute3 \\
        \hline
1 &0.29 &0.32 &0.33 \\
\hline
2 &0.67& 0.32 &0.33\\
\hline
3 &0.67 &0.67 &0.33\\
\hline
4 &0.67& 0.67 &0.76\\
\hline
\end{tabular}
\caption{Position of the real sources ($\cs^*$) for the second synthetic dataset.}
\label{tab: sources tetra}
\end{table}

A method that estimates the sources position from the detected sources is developed for this situation.

\begin{figure}[H]
\centering 
\subfloat[]{\includegraphics[scale=0.2]{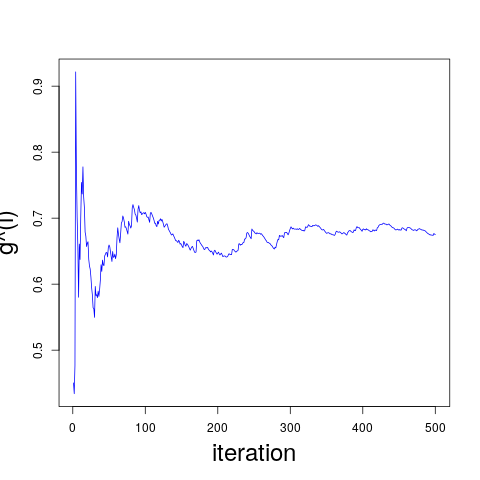}}
\subfloat[]{\includegraphics[scale=0.2]{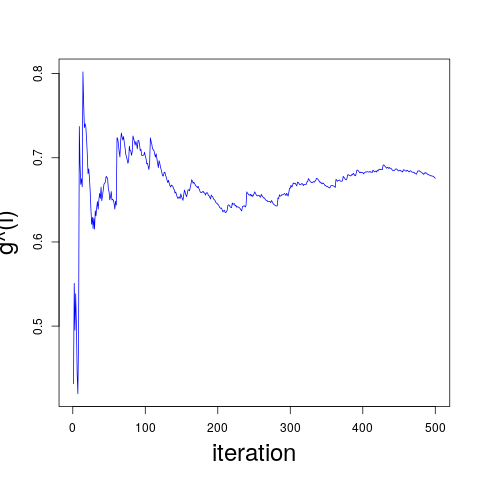}}
\subfloat[]{\includegraphics[scale=0.2]{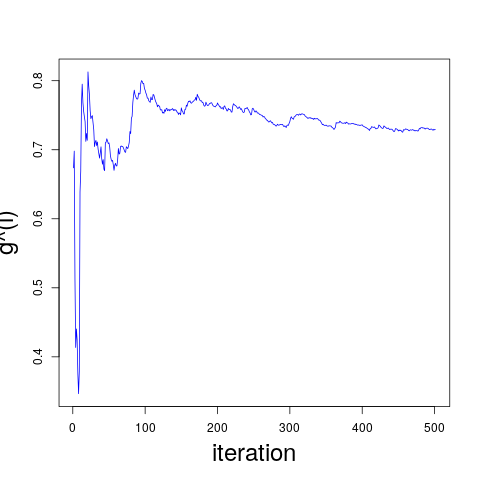}}\\
\subfloat[]{\includegraphics[scale=0.2]{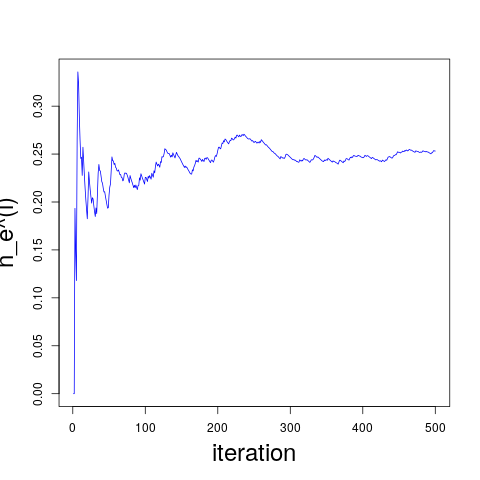}}
\subfloat[]{\includegraphics[scale=0.2]{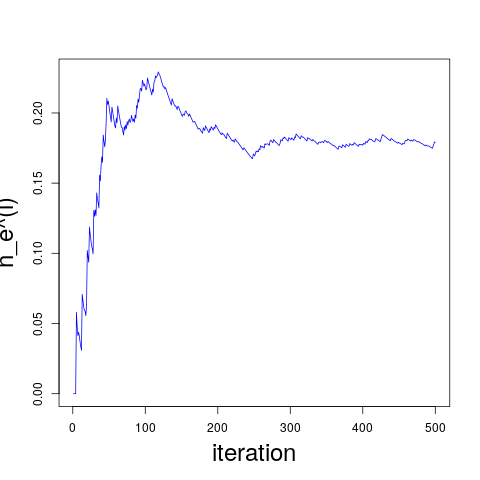}}
\subfloat[]{\includegraphics[scale=0.2]{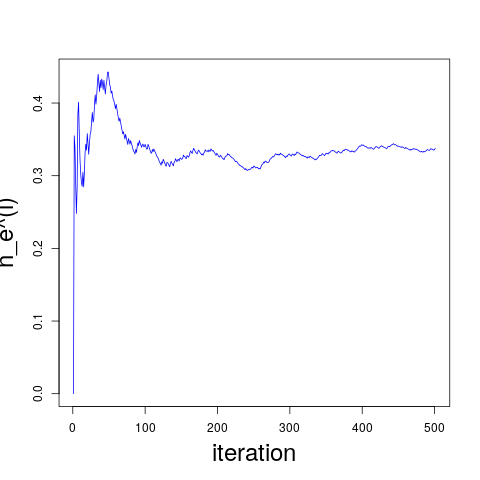}}\\
\subfloat[]{\includegraphics[scale=0.2]{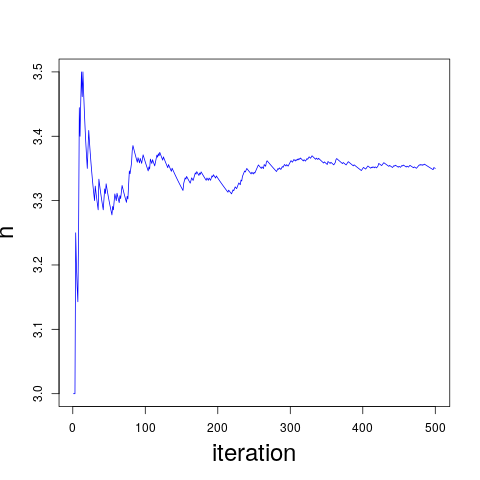}}\\
\subfloat[]{\includegraphics[scale=0.2]{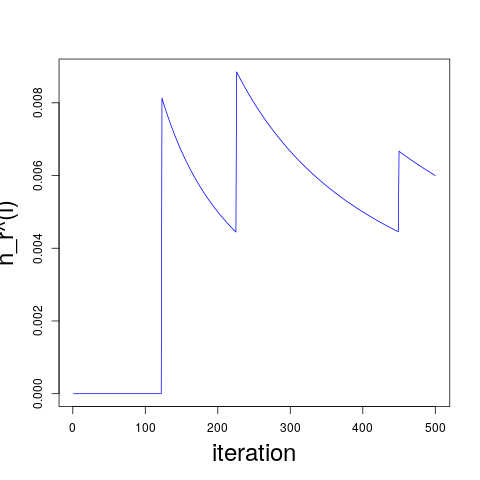}}
\subfloat[]{\includegraphics[scale=0.2]{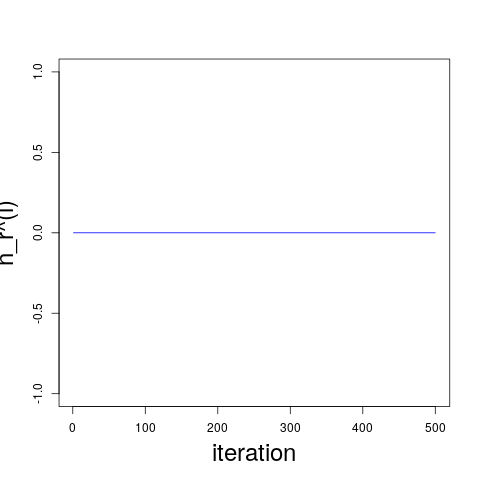}}
\subfloat[]{\includegraphics[scale=0.2]{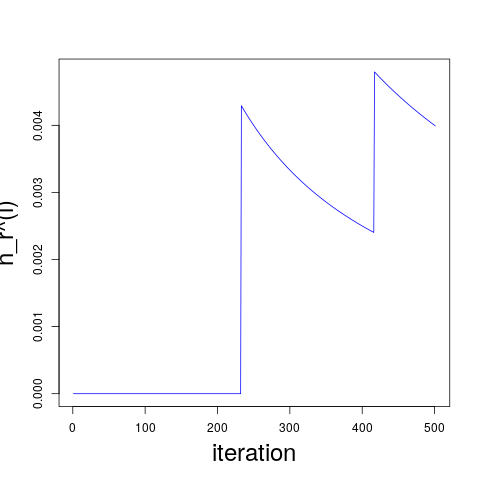}}
\caption{Cumulative means of the statistics for the second synthetic dataset: first plane (a,d,g,h), second plane (b,e,g,i) and third plane (c,f,g,j). Each row represents a statistic.} 
\label{plot: stat tetra} 
\end{figure}

First, the Hug model is applied. The model and dynamics set-up was the same as for the previous dataset. Similarly as before, the cumulative means for the sufficient statistics are computed from the last $500$ saved outputs with the same initialisation as for the previous dataset. The results are shown in Figure~\ref{plot: stat tetra}. It can be noticed that the average number of sources is greater than $3$ in each projected plane.

\begin{figure}[H]
\centering 
\subfloat[first plane]{\includegraphics[scale=0.35]{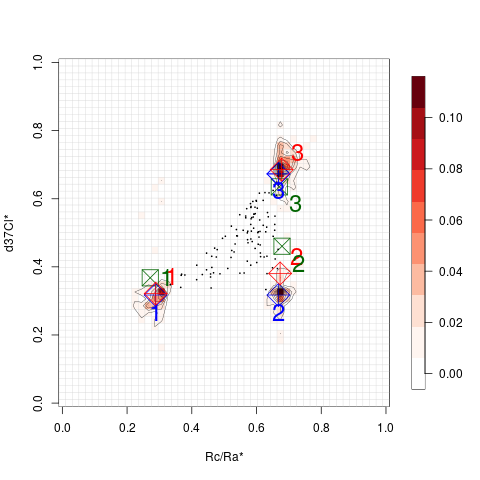}}
\subfloat[second plane]{\includegraphics[scale=0.35]{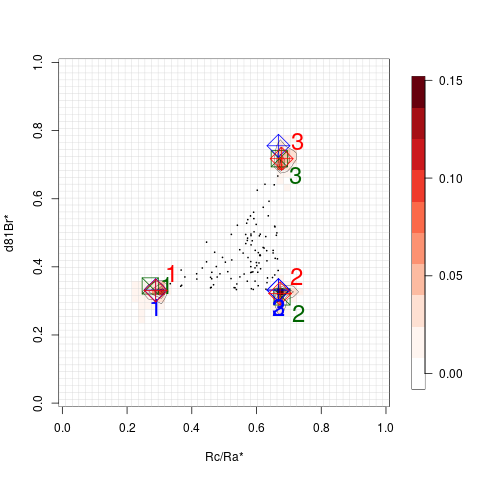}}\\
\subfloat[third plane]{\includegraphics[scale=0.35]{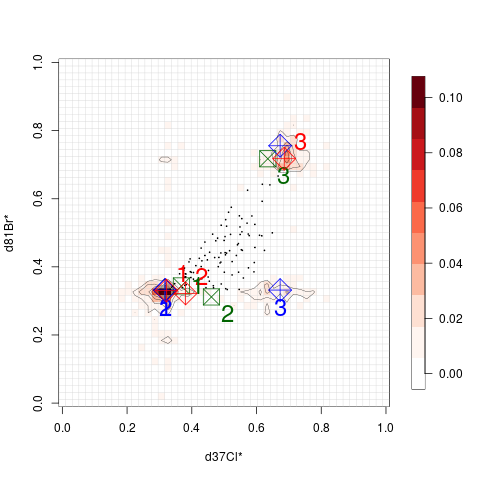}}
\caption{Level sets computed the second synthetic dataset. The blue symbols represent the known sources, the green symbols are the centres of the clusters obtained by the $k$-means algorithm with three clusters on the pattern of sources, and the red symbols are the median points.} 
\label{plot: count tetra} 
\end{figure}

Figure \ref{plot: count tetra} shows the source projections in each plane. The estimated level sets indicate three major regions in each projection plane. The evolution of the third statistics (the number of sources), and more specifically the mean value of sources, does not always allow concluding on the exact number of sources and by extension their position.

The number of sources cannot be known by only considering these statistics. In order to remediate this difficulty, a sequential $k-$means is proposed.

First a projection plane is chosen, the number of classes is chosen depending on the observed value of $n(\cs)$. Next, a $k-$means is performed in the projection plane. Finally, the coordinates of the projected sources are replaced with the coordinates of the detected cluster centres. The algorithm, described in the following, is iterated till convergence, by choosing another remaining projection plane.

\fbox{
\begin{minipage}{\textwidth}
\begin{enumerate}
\item[]\textbf{Algorithm Sequential $k$-means~:} Let $V={1,\dots,L}$ be the set of all the plane  and for each plane $v$, fix the number of clusters $k_v$, the simulated sources $S=(\cs_1,\dots,\cs_n)$. 
\item Till $V=\emptyset$
\begin{itemize}
\item Choose $v$ uniformly without replacement in the set of projection planes
\item Apply $k$-means with $k_v$ clusters on the projection of $S$
\item Replace the coordinate of each simulated sources by the coordinate of the centre of its cluster $c_k^{v}$
\item Remove $v$ in $V$
\end{itemize}
\item Return $S$.
\end{enumerate}
\end{minipage}
}

By applying the sequential $k$-means algorithm with $3$ clusters in each plane, we obtain a pattern with $4$ sources, rather close to the real sources,  described in Table \ref{tab: seq tetra}.

\begin{table}[H]
    \begin{tabular}{|c||c|c|c|c|}
    \hline
        \textbf{Sources} & solute1 & solute2 & solute3 \\
        \hline
1 &0.27  &0.35  &0.33 \\
\hline
2 &0.67  &0.35  &0.33\\
\hline
3 &0.67  &0.64  &0.33\\
\hline
4 &0.67  &0.64  &0.72\\
        \hline
    \end{tabular}
    \caption{Position of the sources obtained from the sequential $k$-means clustering.}
\label{tab: seq tetra}
\end{table}

\begin{figure}[H]
\centering 
\subfloat[first plane]{\includegraphics[scale=0.35]{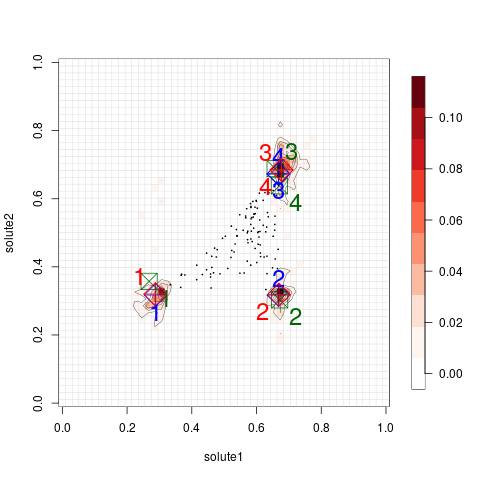}}
\subfloat[second plane]{\includegraphics[scale=0.35]{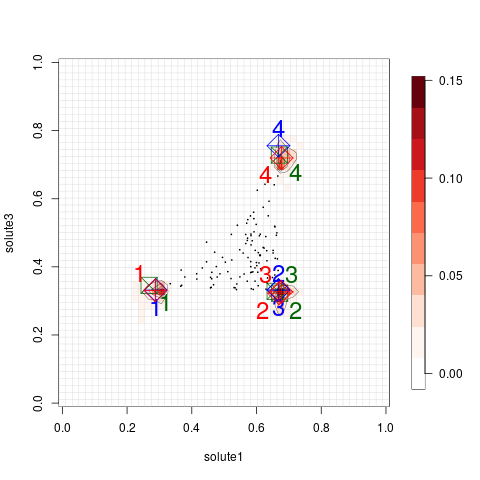}}\\
\subfloat[third plane]{\includegraphics[scale=0.35]{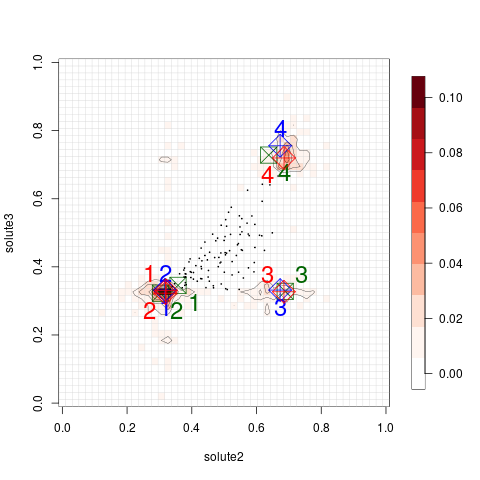}}
\caption{Level sets computed the second synthetic dataset. The blue symbols represent the known sources, the green symbols are the centres of the clusters obtained by the $k$-means algorithm with four clusters, and the red symbols are the median points.} 
\label{plot: count tetra 2} 
\end{figure}

A verification is done by performing a $k$-means with $4$ clusters on the saved pattern in the whole data space. As previously, the proposed pattern is made by the median point of each cluster and is described in the Table \ref{tab: clus tetra} and represented in Figure \ref{plot: count tetra 2}. This pattern appears to be satisfactory: the proposed sources, as numerous as the real sources, are close to the real sources.

\begin{table}[H]
\begin{tabular}{|c||c|c|c|c|c|c|c|c|c|c|}
    \hline
       \multirow{2}{*}{\textbf{Sources}} & \multicolumn{3}{|c|}{solute1} & \multicolumn{3}{|c|}{solute2} & \multicolumn{3}{|c|}{solute3} \\
        \cline{2-10}
        & median & mean & sd& median & mean & sd& median & mean & sd\\
        \hline\hline
1 & 0.68& 0.66& 0.09& 0.68 &0.64 &0.13 &0.71& 0.71& 0.07 \\
\hline
2 &0.67& 0.65& 0.06& 0.32& 0.32& 0.07& 0.33& 0.36& 0.11\\
\hline
3 &0.3& 0.29& 0.06& 0.33& 0.37& 0.12& 0.33& 0.35& 0.1\\
\hline
4 &0.66& 0.63 &0.1& 0.64& 0.63& 0.08& 0.33 &0.35& 0.08\\       
        \hline
    \end{tabular}
    \caption{Proposed pattern for the second synthetic dataset made by the median points of clusters computed using a $k$-means with $4$ classes on the whole space.}
\label{tab: clus tetra}
\end{table}

Clearly, such a procedure is not needed if the exact number of sources is known. But this is precisely the problem to be solved. The previous application validates \textit{a posteriori} the sequential $k$-means algorithm.

The error between the proposed sources and the real sources are given in the Table \ref{tab: error tetra}. The average percentage are all under $5\%$.

\begin{table}[H]
    \begin{tabular}{|c||c|c|c||c|}
    \hline
        \textbf{Sources} & solute1 & solute2 & solute3 & Mean Error Source\\
        \hline
1 &0.0    & 0.0  & 0.0 & 0.0\\
\hline
2 & 0.0   &  0.0   & 3.0 & 1.0\\
\hline
3 &1.5    & 1.5   & 0.0 & 1.0\\
\hline
4 & 1.5    & 1.5   & 5.3 & 2.8\\
\hline\hline
Mean Error Dimension& 0.8  &   0.8  &  2.1   &  1.2\\
        \hline
    \end{tabular}
    \caption{Relative difference (in $\%$) for the proposed pattern of sources with respect to the known ones, and the mean error for each source and each dimension, for the second synthetic dataset.}
\label{tab: error tetra}
\end{table}

\begin{figure}[H]
\centering 
\includegraphics[scale=0.36]{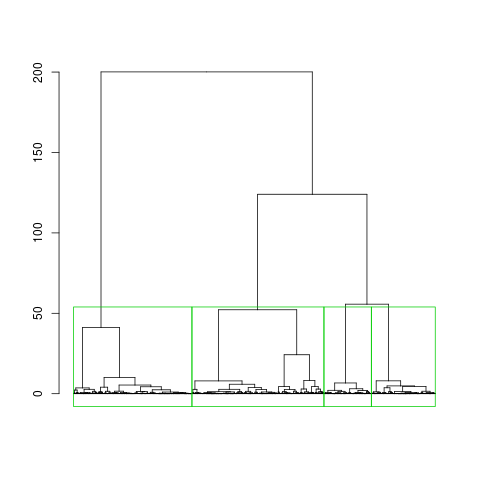}
\caption{Dendrogram obtain from a hierarchical clustering that minimised the within-cluster variance for the second synthetic dataset. Each green rectangle contains the simulated sources that belong to one of the four clusters.} 
\label{plot: dendrogram tetra} 
\end{figure}

The number of sources may also be deduced from the hierarchical clustering algorithm. This clustering is iterative and minimises in each step the within-cluster variance. At each step the cluster, made by merging two existing clusters, with the smallest variance is created. The algorithm ends when only one cluster remains. The within-cluster variance at each step is represented in the dendrogram of Figure \ref{plot: dendrogram tetra}.

As seen in the figure, the within-cluster variance is high if less than $4$ clusters are considered. However, the variance does not decrease significantly when more than $9$ clusters are considered. Hence, the number of sources is assumed to be between $4$ and $9$. The green boxes contain the $4$ clusters.

The saved configurations of sources are now clustered in $5$, $6$, $7$, $8$ and $9$ clusters. The proportion of sources in the four biggest clusters are given in the Table \ref{tab: proba cluster tetra}. In each case, the four biggest clusters contain at least $75\%$ of the sources: the sources can be clustered into four clusters.

\begin{table}[H]
    \begin{tabular}{|m{5em}|c|c|c|c|c|}
    \hline
         & 5 & 6 & 7 & 8 & 9 \\
        \hline
proportion & 0.92 &0.84& 0.82& 0.75& 0.77\\
        \hline
    \end{tabular}
    \caption{Proportion of simulated sources in the $4$ clusters containing the most simulated sources when the $k$-means is applied with $5$, $6$, $7$, $8$ and $9$ clusters.}
\label{tab: proba cluster tetra}
\end{table}

\subsection{Real datasets}

The previously described methods are applied to two real datasets.

\subsubsection{First dataset :} The first real dataset on which the Hug model is applied is from \cite{Pint20}. In this dataset the stable isotopic composition of chlorine $\delta^{37}$Cl, the stable isotopic composition of bromine $\delta^{81}$Br and the stable isotopic composition of helium Rc/Ra ($^3$He$/^4$He normalized to that of the Atmosphere and corrected for the air component) are measured on $m=75$ samples from geothermal wells from Mexico and are supposed to be the result of a three-source mixing system  (mantle, subduction and crust). Note that halogens and noble gases behave conservatively during fluid mixing and that the mixing trends in the two planes considered here are not affected by curvature \cite{Pint20} so that the conditions of the use of the Hug model apply here and $L=2$. 

Because the data are supposed results of a three-source mixing system, the source can be considered as the vertex of a triangle containing the data on each plane. In \cite{Pint20}, the sources are estimated by the vertex of the smallest triangle (in terms of area) containing the data on each plane (Table \ref{tab: pinti}).

\begin{table}[H]
    \begin{tabular}{c|c|c|c|c|c|c|c|}
\cline{2-3} 
\cline{5-6} 
sources &Rc/Ra &$\delta^{37}$Cl in \permille & & Rc/Ra &$\delta^{81}$Br in \permille \\
\cline{2-3} 
\cline{5-6} 
1 (mantle) &7.76 &0.88 & & 8.26 &0.75\\
\cline{2-3} 
\cline{5-6} 
2 (subduction) &6.45 &-0.43& &7.17 &-1.03\\
\cline{2-3} 
\cline{5-6}  
3 (crust) &1.68 &0.11& &1.89 &0.26\\
\cline{2-3} 
\cline{5-6} 
    \end{tabular}
    \caption{Sources estimated in \cite{Pint20} when the data are supposed bi-dimensional and resulting from a three-source mixing system. The sources are the vertex of the smallest triangle that contains the data.}
\label{tab: pinti}
\end{table}

To reconstruct the sources in the $3$ dimensional space, these sources are merged by the coordinate that they have in common: their value is the means between the two previous values.
The reconstructed sources are described in Table \ref{tab: reconstructed pinti}.

\begin{table}[H]
    \begin{tabular}{c|c|c|c|c|c|c|c|}
\cline{2-4} 
sources &Rc/Ra &$\delta^{37}$Cl in \permille  & $\delta^{81}$Br in \permille \\
\cline{2-4}
1 (mantle) &8.01 &0.88  &0.75\\
\cline{2-4} 
2 (subduction) &6.81 &-0.43 &-1.03\\
\cline{2-4}  
3 (crust) &1.78  &0.11 & 0.26\\
\cline{2-4} 
    \end{tabular}
    \caption{Composition of the sources estimated in \cite{Pint20} in the complete data space. The reconstruction is made by merging by the coordinate Rc/Ra. The value of this coordinate is the means between this coordinate on each plane.}
\label{tab: reconstructed pinti}
\end{table}

\begin{figure}[H]
\centering 
\subfloat[first normalised plane ]{\includegraphics[scale=0.35]{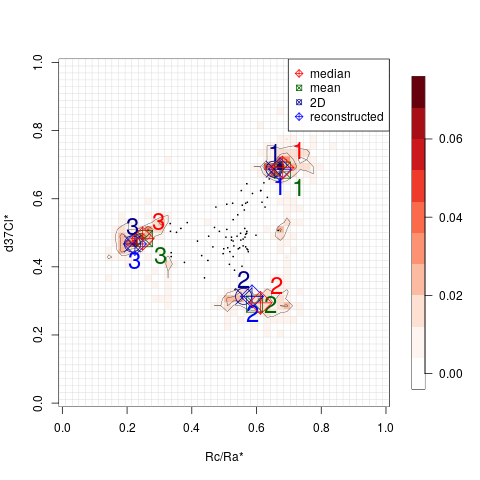}} 
\subfloat[second normalised plane ]{\includegraphics[scale=0.35]{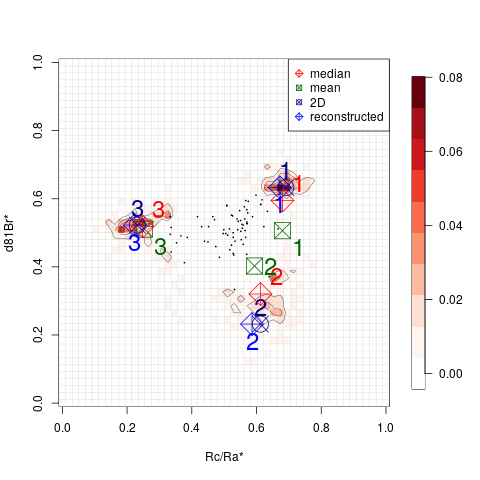}} 
\caption{Level sets computed the first real dataset~\cite{Pint20}. The blue symbols represent the sources reconstructed by \cite{Pint20}, the green symbols are the centres of the clusters obtained by the $k$-means algorithm with three clusters on the pattern of sources, and the red symbols are the median points.} 
\label{plot: level Pinti} 
\end{figure}

The Hug model is applied with the same initialisation as in the previous section. As previously, the last $500$ saved configurations are projected on every normalised plane in regular grid with cells of length $0.02$. The normalised dimensions are indicated by adding a $*$ to the raw dimensions. Figure \ref{plot: level Pinti} and Table \ref{tab: proposed pinti} present the results. The blue symbols are the previously bi-dimensional mentioned sources and the reconstructed sources. The model finds on each plane $3$ areas with high probability of containing an estimated source. By applying the sequential $k$-means algorithm with $3$ clusters on each plane, $3$ sources are estimated. These sources are given by the $k$-means algorithm with $3$ clusters. 

The proposed pattern of sources is not useful in this state because it can not be compared to the raw data and the sources of Table \ref{tab: pinti}. Hence, the reverse procedure of the normalisation, presented in section \ref{transformation}, has to be applied on the detected sources and the proposed pattern in Table \ref{tab: proposed pinti true}.

\begin{figure}[H]
\centering 
\subfloat[first plane]{\includegraphics[scale=0.35]{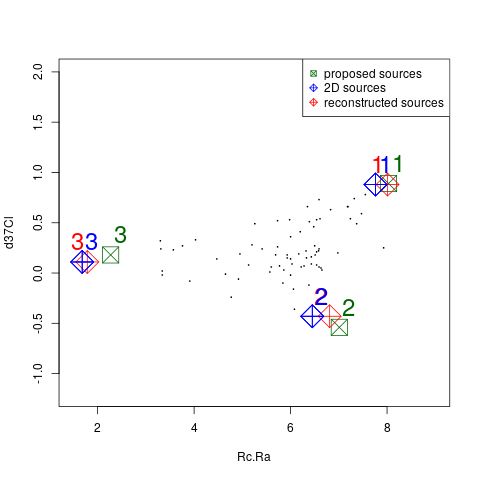}} 
\subfloat[second plane]{\includegraphics[scale=0.35]{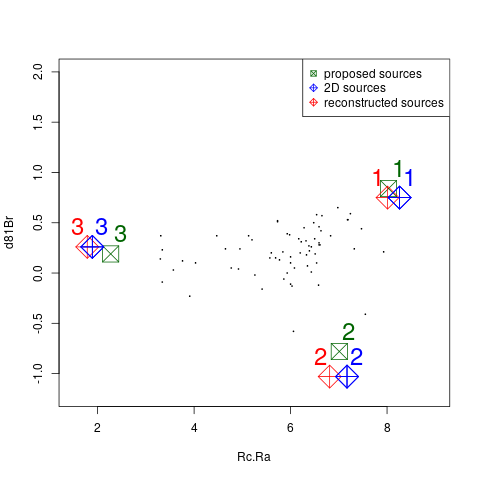}} 
\caption{Composition of the sources proposed by the Hug model in the initial data space. The composition in $\delta^{37}$Cl and $\delta^{81}$Br are given in \permille.} 
\label{plot: true Pinti} 
\end{figure} 

\begin{table}[H]
    \begin{tabular}{|c||c|c|c|c|c|c|c|c|c|c|}
    \hline
       \multirow{2}{*}{\textbf{Sources}} & \multicolumn{3}{|c|}{Rc/Ra} & \multicolumn{3}{|c|}{$\delta^{37}$Cl} & \multicolumn{3}{|c|}{ $\delta^{81}$Br} \\
        \cline{2-10}
        & median & mean & sd& median & mean & sd& median & mean & sd\\
        \hline\hline
1 (mantle) &0.69 &0.69 &0.06 &0.71 &0.7 &0.07 &0.64 &0.59 &0.13\\
\hline
2 (subduction) &0.6& 0.6& 0.08 &0.29& 0.29& 0.07& 0.26& 0.3& 0.14\\
\hline
3 (crust) &0.23& 0.24 &0.07& 0.48& 0.47& 0.06& 0.52& 0.52 &0.06\\
\hline
    \end{tabular}   
    \caption{Proposed pattern for the first real dataset, made by clustering the sources in three clusters  computed using a $k$-means algorithm, on the whole normalised data space.}
    \label{tab: proposed pinti}
\end{table}

\begin{table}[H]
    \begin{tabular}{c|c|c|c|c|c|c|c|}
\cline{2-4} 
sources &Rc/Ra &$\delta^{37}$Cl in \permille & $\delta^{81}$Br in \permille \\
\cline{2-4}
1 (mantle) &8.12  &0.90 & 0.58\\
\cline{2-4} 
2 (subduction) &7.17 &-0.50& -0.64\\
\cline{2-4}  
3 (crust) &2.12 & 0.17&  0.24\\
\cline{2-4} 
    \end{tabular}  
    \caption{Proposed pattern for the first real dataset made by the median points of clusters computed using a $k$-means with $3$ classes on the initial data space (after reversing the normalisation).}
\label{tab: proposed pinti true}
\end{table}

The relative difference (in $\%$) between the reconstructed sources and the proposed sources, after reversing the transformation, are given in Table \ref{tab: error pinti}.

\begin{table}[H]
    \begin{tabular}{|c||c|c|c||c|}
    \hline
        \textbf{Sources} & Rc/Ra &$\delta^{37}$Cl & $\delta^{81}$Br & Mean Error Source\\
        \hline
1 &1.4  &2.3 &22.7 & 8.8\\
\hline
2 & 5.3& 16.3 &37.9 & 19.8\\
\hline
3 &19.1 &54.5  &7.7 & 27.1\\
\hline\hline
Mean Error Dimension&8.6 & 24.4 &22.7    &  18.6\\
        \hline
    \end{tabular}
    \caption{Relative difference (in $\%$) for the detected sources with respect to the sources estimated in \cite{Pint20}, and the mean error for each source and each dimension, for the first real dataset.}
\label{tab: error pinti}
\end{table}

Eventually, applying the Hug model provides fairly consistent results with the geometrical approach proposed by \cite{Pint20} on this rather simple dataset (i.e. $3$ parameters considered and $3$ proposed sources). The relatively small differences in composition of the sources detected by the Hug model compared to those proposed by \cite{Pint20} do not imply to reconsider the geological interpretations regarding the origin of the geothermal fluids. As seen previously, working only on planes may induce bias. Thus, in \cite{Pint20} at least $3$ sources are detected (more sources may be needed) whereas the Hug model detect exactly $3$ sources.

\subsubsection{Second dataset :}

\begin{figure}[H]
\centering 
\subfloat[normalised plane $1$]{\includegraphics[scale=0.35]{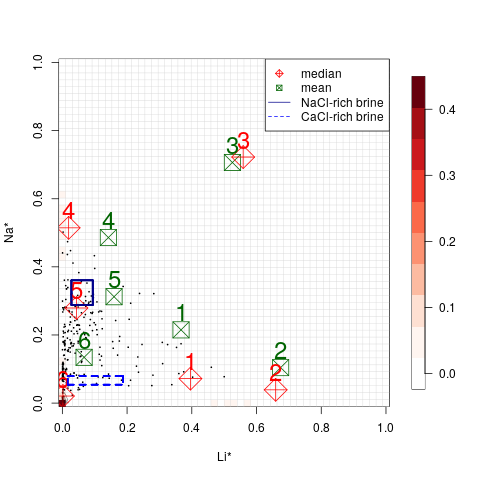}} 
\subfloat[normalised plane $2$]{\includegraphics[scale=0.35]{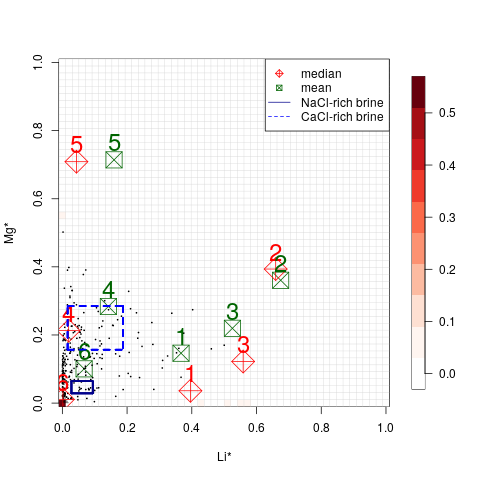}}\\
\subfloat[normalised plane $3$]{\includegraphics[scale=0.35]{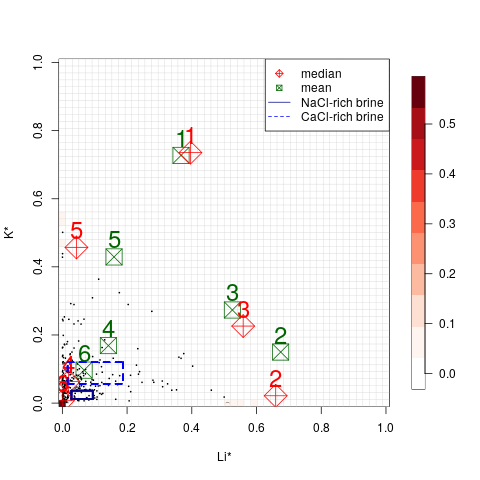}}
\subfloat[normalised plane $4$]{\includegraphics[scale=0.35]{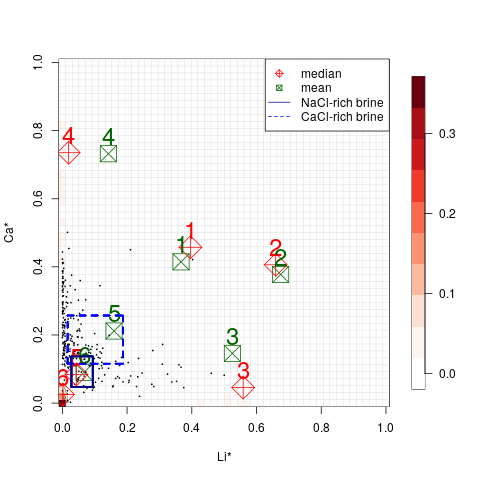}} 
\caption{Level sets computed the second real dataset. The blue rectangles made by the continuous line and the dotted line represent respectively the NaCl-rich brine and the Ca$Cl_2$-rich brine presented in \cite{RicCat15}.} 
\label{plot: count Athabasca} 
\end{figure}

\begin{figure}[H]
\centering 
\subfloat[normalised plane $5$]{\includegraphics[scale=0.35]{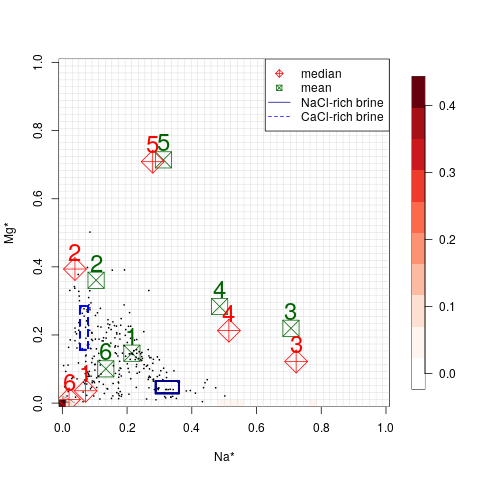}} 
\subfloat[normalised plane $6$]{\includegraphics[scale=0.35]{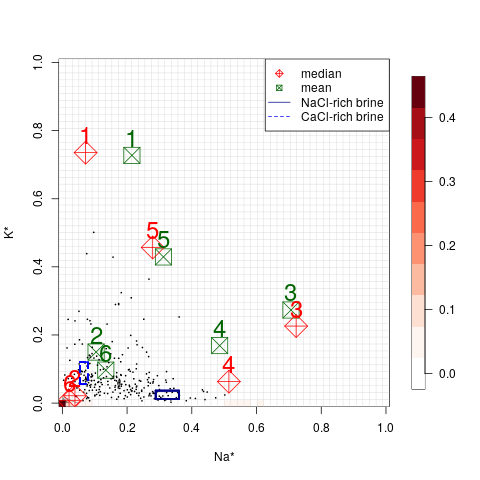}}\\
\subfloat[normalised plane $7$]{\includegraphics[scale=0.35]{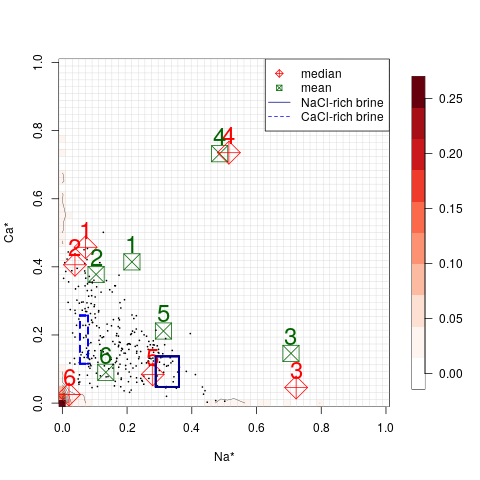}}  
\subfloat[normalised plane $8$]{\includegraphics[scale=0.35]{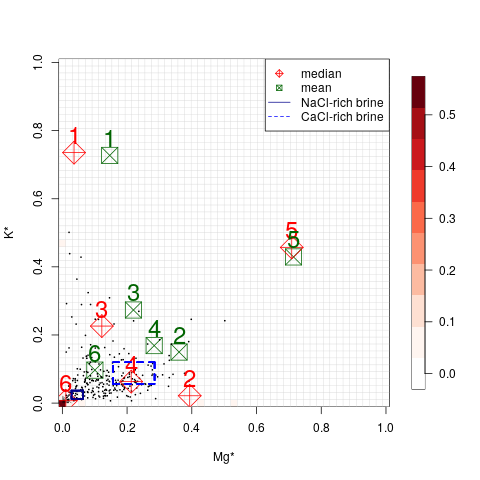}}\\
\subfloat[normalised plane $9$]{\includegraphics[scale=0.35]{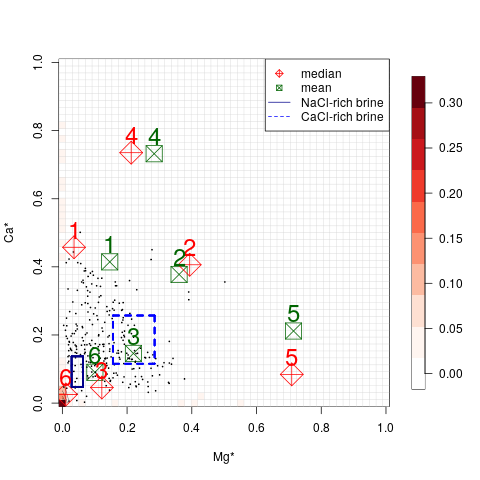}}
\subfloat[normalised plane $10$]{\includegraphics[scale=0.35]{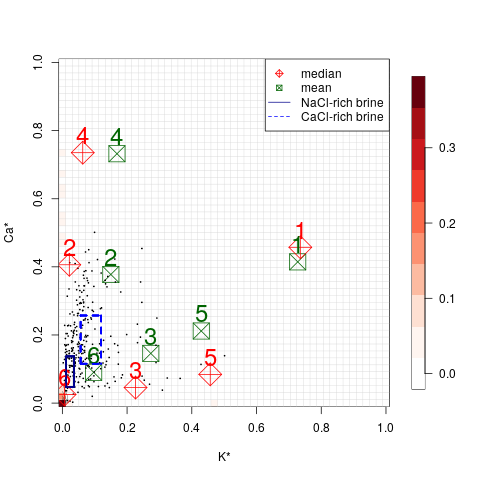}} 
\caption{Level sets computed the second real dataset. The blue rectangles made by the continuous line and the dotted line represent respectively the NaCl-rich brine and the Ca$Cl_2$-rich brine presented in \cite{RicCat15}.} 
\label{plot: count Athabasca2} 
\end{figure}

The second dataset considered is from \cite{RicPet10}, \cite{RicCat15} and \cite{MarMer19}.
The dataset accounts for the composition of fluid inclusions from a series of hydrothermal uranium deposits of the Athabasca Basin (Canada). The concentrations of five chemical elements (lithium Li, sodium Na, magnesium Mg, potassium K and calcium Ca) are obtained by Laser Ablation-Inductively Coupled Plasma Mass Spectrometry (LA-ICPMS). Based on graphical interpretation of two-dimentional composition diagrams, previous authors have concluded that the data are spread between a "NaCl-rich-brine end-member" and a "Ca$Cl_2$-rich brine endmember" where NaCl-rich brines are definied by all fluid inclusions with [Na] $> 80000$ ppm and Ca$Cl_2$-rich brines are defined as any fluid inclusions with [Na] $< 30000$ ppm. The main conclusion was that the data result for a mixing of two sources. The full composition of the NaCl-rich and Ca$Cl_2$-rich brine end-members
are given in Table \ref{tab: true sources}.
The last $500$ saved pattern are projected on the $10$ normalised planes in regular grid with cells of length $0.02$. The results are shown in the Figures \ref{plot: count Athabasca} and \ref{plot: count Athabasca2}. On each plane, the model detects $3$ areas with a rather high probability, indicating the presence of potential sources. By applying the sequential $k$-means with $3$ clusters on each plane, $6$ sources are detected.

\begin{table}[H]
    \begin{tabular}{|c||c|c|c|c|}
    \hline
\multirow{2}{*}{\textbf{Sources}} &\multicolumn{2}{|c|}{NaCl-rich brine}&\multicolumn{2}{|c|}{$CaCl_2$-rich brine} \\
        \cline{2-5}
        & Q25 & Q75 &Q25&Q75\\
        \hline\hline
[Li] in ppm &900&3000&520&6000\\
        \hline
[Na] in ppm &80000&100000&15000&22000\\
        \hline
[Mg] in ppm  & 4000&  9000& 22000  &   40000  \\
        \hline
[K] in ppm &     1700&5200&8000&17000\\
        \hline
[Ca] in ppm & 11000&32000&27000&60000 \\
        \hline
    \end{tabular}
    \caption{Range of the sources detected in \cite{RicCat15}. The data are regroup in a group containing the data with $[Na]>80000$ and a group containing data with $[Na]<30000$. For these groups, respectively NaCl-rich brine and Ca$Cl_2$-rich brine, are given the quantile at $25\%$ (Q25) and $75\%$ (Q75).} 
\label{tab: true sources}
\end{table}

\begin{figure}[H]
\centering 
\includegraphics[scale=0.36]{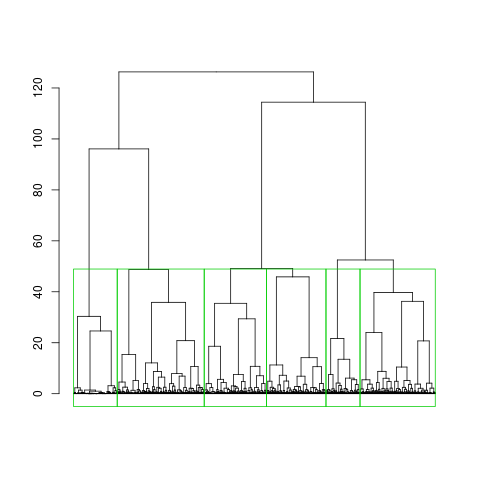}
\caption{Dendrogram obtain from a hierarchical clustering that minimised the within-cluster variance. Each green rectangle contains the simulated sources that belong to one of the six clusters.} 
\label{plot: dendrogram Athab} 
\end{figure}

The hierarchical cluster algorithm is also applied to confirm the number of sources. As seen in Figure \ref{plot: dendrogram Athab}, the within-cluster variance is rather high when less than $6$ clusters are considered. Moreover, when more than $10$ clusters are considered, the variance is very low. In conclusion, the number of clusters should be between $6$ and $9$. 

The proportion of sources in the $6$ biggest clusters obtained by a $k$-means with $7$, $8$ and $9$ clusters, is given in Table \ref{tab: proba cluster Athab}. The proportion is greater than $70\%$, hence the hypothesis of $6$ clusters can be reasonably confirmed.

\begin{table}[H]
    \begin{tabular}{|m{5em}|c|c|c|}
    \hline
         &  7 & 8 & 9 \\
        \hline
proportion  & 0.90 &0.79& 0.72\\
        \hline
    \end{tabular}
    \caption{Proportion of sources in the $6$ clusters containing the most simulated sources when the $k$-means is applied with $7$, $8$ and $9$ clusters.}
\label{tab: proba cluster Athab}
\end{table}

The proposed pattern of sources is given in Table \ref{tab: proposed sources}.

\begin{table}[H]
    \begin{tabular}{|c|*{15}{|l}|}
    \hline
\multirow{2}{*}{} & \multicolumn{3}{|c|}{Li*} & \multicolumn{3}{|c|}{Na*} & \multicolumn{3}{|c|}{Mg*}& \multicolumn{3}{|c|}{K*}& \multicolumn{3}{|c|}{Ca*}\\
\cline{2-16}
        & med & mean & sd& med & mean & sd& med & mean & sd& med & mean & sd& med & mean & sd\\
        \hline\hline
1&0&0.15&0.21&0.03&0.15&0.2&0.01&0.08&0.14&0.55&0.55&0.14&0.03&0.12&0.18\\
        \hline
2&0&0.05&0.12&0.02&0.1&0.17&0.59&0.59&0.14&0.02&0.18&0.23&0.04&0.15&0.2\\
        \hline
3&0&0.09&0.18&0.55&0.55&0.19&0.01&0.09&0.16&0&0.08&0.15&0.01&0.04&0.06\\
        \hline
4&0&0.15&0.22&0.01&0.04&0.07&0.01&0.13&0.19&0.01&0.11&0.18&0.63&0.64&0.15\\
        \hline
5&0.57&0.61&0.15&0.05&0.15&0.2&0.07&0.22&0.24&0.01&0.09&0.16&0.02&0.1&0.16\\
        \hline
6&0&0.1&0.2&0.56&0.58&0.15&0.02&0.17&0.23&0.02&0.16&0.24&0.61&0.63&0.16\\
        \hline
    \end{tabular}
    \caption{Proposed pattern for the second real dataset, made by clustering the sources in six clusters  computed using a $k$-means algorithm, on the whole normalised data space.}
    \label{tab: proposed sources}
\end{table}

\begin{figure}[H]
\centering 
\subfloat{\includegraphics[scale=0.35]{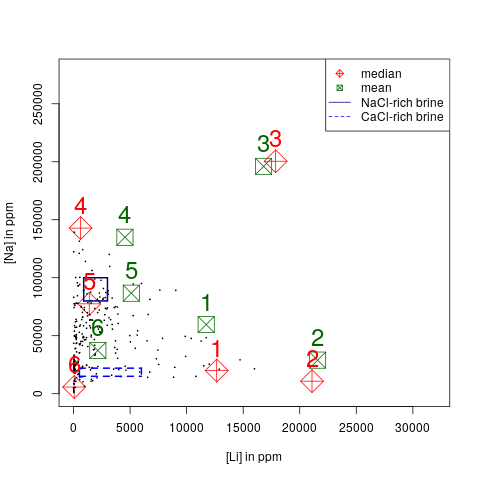}} 
\subfloat{\includegraphics[scale=0.35]{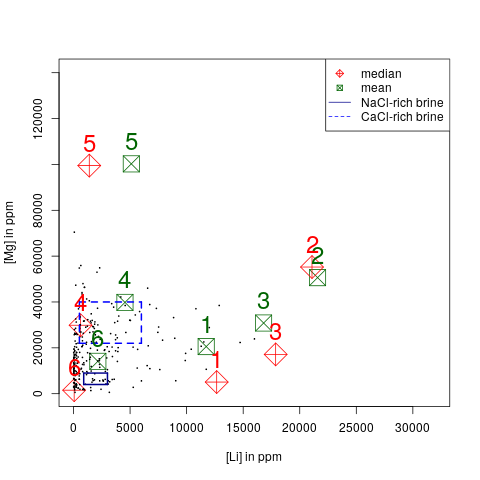}}\\
\subfloat{\includegraphics[scale=0.35]{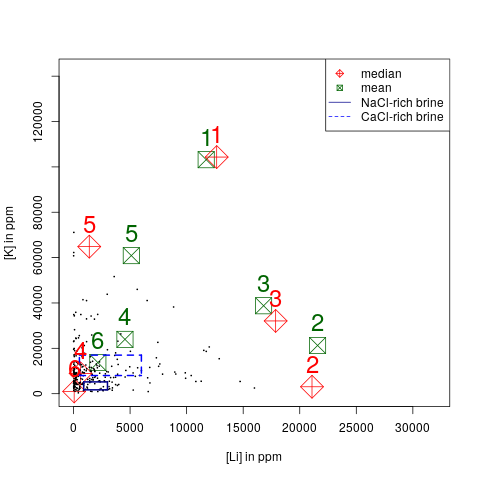}}
\subfloat{\includegraphics[scale=0.35]{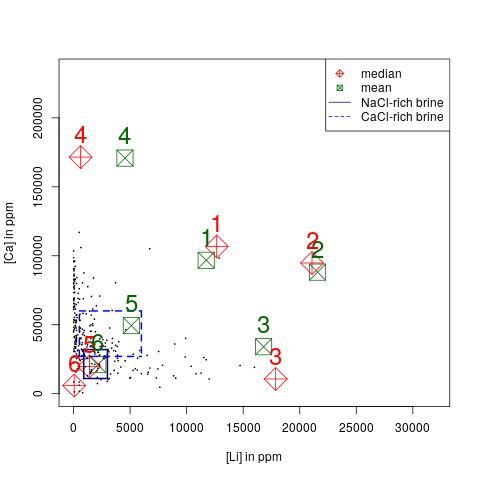}}
\caption{Composition of the proposed sources for the second real dataset, made by clustering the sources in six clusters computed using a $k$-means algorithm, on the first four planes (after reversing the normalisation). The blue rectangle and the dotted rectangle represent respectively the NaCl-rich brine and the Ca$Cl_2$-rich brine presented in \cite{RicCat15}.} 
\label{plot: true Athabasca} 
\end{figure}

\begin{figure}[H]
\centering 
\subfloat{\includegraphics[scale=0.35]{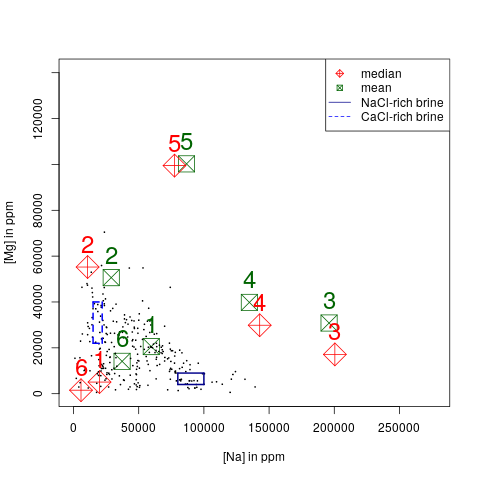}} 
\subfloat{\includegraphics[scale=0.35]{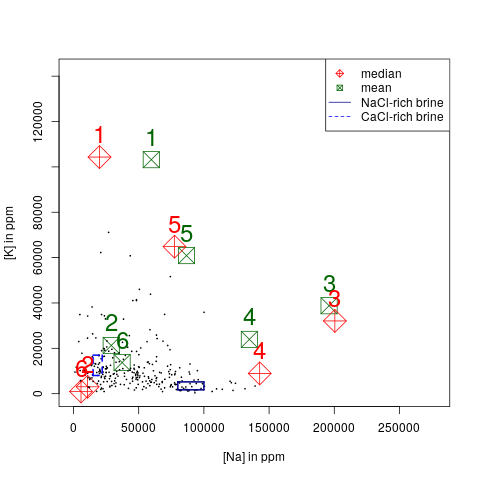}}\\
\subfloat{\includegraphics[scale=0.35]{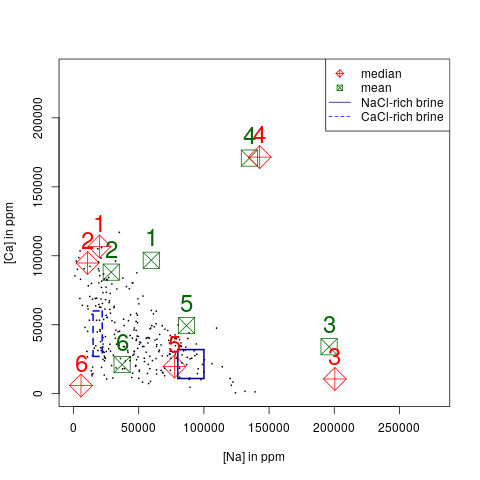}}  
\subfloat{\includegraphics[scale=0.35]{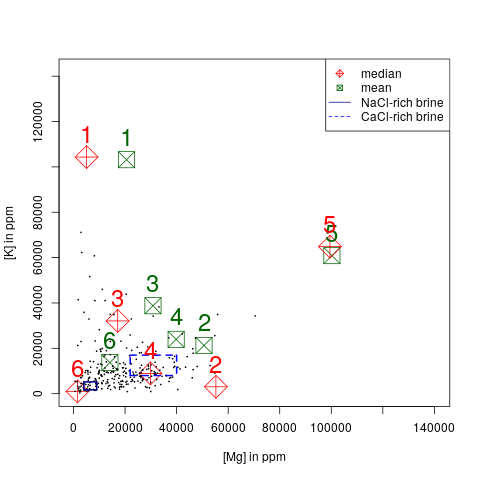}} \\
\subfloat{\includegraphics[scale=0.35]{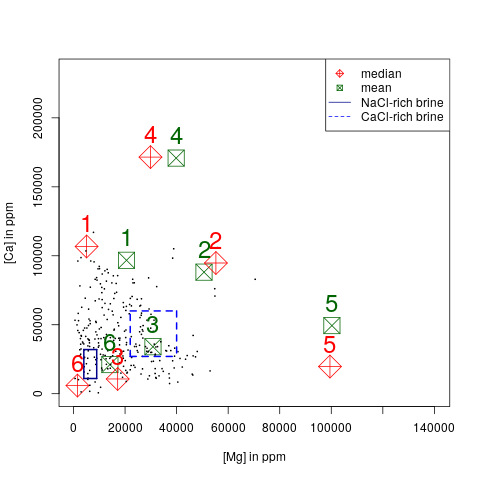}} 
\subfloat{\includegraphics[scale=0.35]{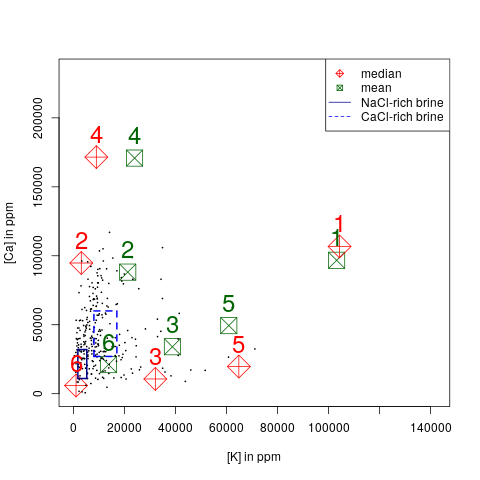}} 
\caption{Composition of the proposed sources for the second real dataset, made by clustering the sources in six clusters computed using a $k$-means algorithm, on the last six planes (after reversing the normalisation). The blue rectangle and the dotted rectangle represent respectively the NaCl-rich brine and the Ca$Cl_2$-rich brine presented in \cite{RicCat15}.} 
\label{plot: true Athabasca2} 
\end{figure}

The proposed pattern of sources is presented in the original data space in the Table \ref{tab: proposed true sources} and the Figures \ref{plot: true Athabasca} and \ref{plot: true Athabasca2}.

On each plane the areas are: a first area (low abscissa values, low ordinate values), a second area  (low abscissa values, high ordinate values) and a last area (high abscissa values, low ordinate values).

\begin{table}[H]
    \begin{tabular}{|c||c|c|c|c|c|}
    \hline

\textbf{Sources} &[Li] in ppm &[Na] in ppm &[Mg] in ppm &[K] in ppm &[Ca] in ppm \\
        \hline
1&12658 & 19934 & 5070& 104325 &106728\\
        \hline
2&21084 & 10813& 55239 &  3051&  94757\\
        \hline
3& 17870 &200441& 17131 & 32081 & 10663\\
        \hline
4&  624 &142790& 29854  & 8944 &171537\\
        \hline
5& 1393 & 77471& 99486 & 64833&  19675\\
        \hline
6&   57 &  5744 & 1524 & 992  & 5900\\
        \hline
    \end{tabular}
    \caption{Composition of the proposed sources, made by clustering the sources in six clusters computed using a $k$-means algorithm, on the whole data space (after reversing the normalisation).}
    \label{tab: proposed true sources}
\end{table}

In \cite{RicPet10}, \cite{RicCat15} and \cite{MarMer19}, the data are supposed the result of a mixing between two brine sources. The Hug model propose six sources. None of the six sources proposed by the HUG model matches consistently the composition of the NaCl-rich or the Ca$Cl_2$-rich brine end-members. The first explanation is that the HUG model detects sources that are close but outside the convex hull of the data, while the NaCl-rich and Ca$Cl_2$-rich brine represent composition end-members. The second explanation is that the HUG model uses consistent determination of sources across all considered elemental compositions (five dimensions and ten planes) while previous graphical detection used only four planes independently (Li versus Na; Ca versus Na; K versus Na and Mg versus Na).   


As the HUG model detects sources without projections biases, one should reconsider the previous interpretations of the composition of fluid inclusions in the Athabasca basin. While the continuum of data is still compatible with mixing of different sources: the nature of the sources detected with HUG can be questioned: some of the detected sources may actually be linked to compositional data that suffered from perfectible analytical quality and therefore may be only considered as simple artefact. The HUG model can therefore be also used for detection of exotic data. In any case, the nomenclature and composition of the NaCl and Ca$Cl_2$-rich brine end-members must be reconsidered. In order to make the difference between the detected source that are real sources and the detected sources that result from exotic data, one should reconsider raw analytical data, which is beyond the scope of this manuscript.

\section{Conclusions and perspectives}
\label{sec:conclusion}
This paper presents a new interaction point process that integrates geological knowledge for the purpose of automatic sources detection. The construction of the model takes into account the multidimensional nature of the data. A Metropolis Hastings within Gibbs simulation dynamics was built for the model in order to manage the multidimensional aspect of the problem.
The source pattern is estimated by the point process configuration that maximise the probability density describing the model. Based on the proposed Metropolis-Hastings dynamics, a simulated annealing algorithm was made, in order to avoid local minima. Level sets estimation is used in order to provide more reliable results and to reduce uncertainties. The adopted strategy to cope with the multidimensionality of the problem, was to perform inference on projection planes. The synthesis of the obtained results was done by constructing a new sequential $k-$means algorithm.


The model parameters set-up was done by using synthetic data where the sources were known. This allowed the construction of parametric priors $p(\theta)$. Detection errors are provided for the considered synthetic datasets. Numerical experiences done using known real datasets already show that the results obtained with our automatic method match the ones presented in the literature. 


Clearly, the prior choice is a crucial point that influences the general performances of the method. Currently, the sensitivity of the model to the prior and the quality of the data is studied. New procedures for inferring the model parameters are also studied in order to improve the quality of the results furnished by the proposed algorithms.


The Hug model is a flexible tool that detects sources patterns in multidimensional data, without knowledge on the number of sources. At our best knowledge, automatic result validation is still an open problem. For the moment, the results should always be confirmed and confronted by an expert.


If it is to list challenges regarding the present approach, we would like to mention the consideration of chemical reactions and curvature effects. Moreover, in this paper the proposed sources are supposed to contribute to every data point. Hence, it may be interesting to introduce consideration of time by searching for sub-systems of mixing or adding the geographic position of the data points. Currently, a thorough study concerning data uncertainties using Bayesian inference is under development.


\section*{Acknowledgements}

This work was performed in the frame of the DEEPSURF project (http://lue.univ-lorraine.fr/fr/impact-deepsurf) at Universit\'e de Lorraine. This work was supported partly by the French PIA project Lorraine Universit\'e d'Excellence, reference ANR-15-IDEX-04-LUE.

\bibliographystyle{apalike}
\bibliography{bibliographie}

\newpage

\appendix 
\pagenumbering{roman}

\section{Appendix: proof}

\label{secA1}

\begin{proof}
Let a configuration $\cs\in\Omega$. By hypothesis there is $M_1\in\RR$ such as
\begin{equation}
M_1=\theta_1\min_{t\in[0,\nu(W)]}\left\{\left\lvert\frac{t}{g^{(l)}(\dd)}-1 \right\rvert\right\}.
\end{equation}
Hence
\begin{equation}
 U^{(l)}(\cs|\theta,\dd) \geq M_1 +\theta_3 n(\cs)+\theta_4 n_{r}^{(l)}(\cs)\geq \theta_3 n(\cs)+\theta_4 n_{r}^{(l)}(\cs).
\end{equation}

The energy function of the Hug model with $K=2$ dominates the energy function of a Strauss point process. A Strauss point process is integrable if $\theta_4 > 0$ \cite{Geye99,Lies00,MollWaag04}. For $\cs \in \Omega$ and  $\xi \in  W$ such as $s_i \neq \xi, \forall s_i \in \cs$, its Papangelou conditional intensity is dominated by
\begin{equation}
\lambda^*(\cs,\xi)=\exp\left[-\theta_3\right]\exp\left[-\theta_4\right]^{n_{r}(\cs\cup \xi)-n_{r}(\cs)}\leq \exp\left[-\theta_3\right].
\end{equation}

The Hug  model with $K=2$ is integrable if $\theta_4 > 0$.
\end{proof}

\end{document}